\documentclass[12pt]{article}
\newcommand{\be}{\begin{equation}}
\newcommand{\ee}{\end{equation}}

\newcommand{\bi}{\begin{itemize}}
\newcommand{\ei}{\end{itemize}}
\newcommand{\bea}{\begin{eqnarray}}
\newcommand{\eea}{\end{eqnarray}}
\newcommand{\ba}{\begin{array}}
\newcommand{\ea}{\end{array}}

\usepackage[left=2.50cm, right=2.50cm, top=2.50cm, bottom=2.50cm]{geometry}
\usepackage[utf8]{inputenc}
\usepackage{cancel}
\usepackage[english]{babel}
\usepackage{physics}
\usepackage{hyperref}
\usepackage{amsthm}
\usepackage{mathrsfs}
\usepackage{mathtools}
\usepackage{graphicx}
\usepackage{changepage}
\usepackage{amssymb,amsmath}
\usepackage{verbatim}
\usepackage{empheq}
\usepackage{xcolor}
\usepackage{tikz}
\usepackage{float}
\usepackage{multirow}
\usepackage{multicol}
\usepackage{amsmath}
\usepackage{centernot}
\usepackage[hang,flushmargin]{footmisc} 
\usepackage[normalem]{ulem}
\usetikzlibrary{tikzmark}
\numberwithin{equation}{section}
\hypersetup{hidelinks}

\newlength{\bibitemsep}
\newlength{\bibparskip}
\let\oldthebibliography\thebibliography
\renewcommand\thebibliography[1]{%
  \oldthebibliography{#1}%
  \setlength{\parskip}{\bibitemsep}%
  \setlength{\itemsep}{\bibparskip}%
}
\begin{document}
\par
\bigskip
\Large
\noindent
{\bf 
Quasi-topological fractons: a 3D dipolar gauge theory\\
\par
\rm
\normalsize

\hrule

\vspace{1cm}

\large
\noindent
{\bf Erica Bertolini$^{1,a}$},
{\bf Alberto Blasi$^{2,b}$}, 
{\bf Nicola Maggiore$^{2,3,c}$}\\

\par

\small

\noindent$^1$ School of Theoretical Physics, Dublin Institute for Advanced Studies, 10 Burlington Road, Dublin 4, Ireland.

\noindent$^2$ Istituto Nazionale di Fisica Nucleare - Sezione di Genova, Via Dodecaneso 33, I-16146 Genova, Italy.

\noindent$^3$ Dipartimento di Fisica, Universit\`a di Genova, Via Dodecaneso 33, I-16146 Genova, Italy.

\smallskip

\vspace{1cm}

\noindent
{\tt Abstract}\\
We consider the theory of a generic rank-2 tensor field in three spacetime dimensions, which involves a symmetric tensor field transforming under infinitesimal diffeomorphisms, and a vector field, whose gauge transformation depends on a local vector parameter. The gauge fixing shows a non-trivial structure, and some non-intuitive possibilities are listed. Despite the fact that the theory is not topological, the energy-momentum tensor vanishes on-shell, which justifies the ``quasi-topological'' appellation we give to this theory.
We show that the theory has three degrees of freedom. Moreover we find an interesting physical interpretation, which consists in a generalized planar electromagnetism and in the emergence of two vector charges with restricted mobility. These are typical fractonic behaviours which can be related to the so called traceless scalar and vector charge theories.


\vspace{\fill}

\noindent{\tt Keywords:} \\
Quantum field theory, tensor gauge field theory, fractons.

\vspace{1cm}

\hrule
\noindent{\tt E-mail:
$^a$ebertolini@stp.dias.ie,
$^b$alberto.blasi@ge.infn.it,
$^c$nicola.maggiore@ge.infn.it.
}
\newpage

\section{Introduction}

Covariance principle is an extremely powerful tool in constructing quantum field theories, General Relativity being the 
most representative example. Once the number of spacetime dimensions, the field content of the theory and the symmetry transformation have been chosen, covariance, together with locality and power counting, determines the corresponding quantum field theory. Limiting ourselves to gauge theories, some examples are
listed in Table 1. 
\begin{table}[H]
\centering
	\begin{tabular}{|c|c|c|c|}
	\hline
	{\bf dimensions}&{\bf field content}&{\bf symmetry} &{\bf theory} \\\hline
4D&$h_{\mu\nu}$&$\delta h_{\mu\nu}=\partial_\mu\xi_\nu +\partial_\nu\xi_\mu $&Linearized Gravity\\\hline
	4D&$A_\mu$&$\delta A_\mu=\partial_\mu\Lambda$&Maxwell\\\hline
	3D&$A_\mu$&$\delta A_\mu=\partial_\mu\Lambda$&Chern-Simons\\\hline
	\end{tabular}
\caption{\footnotesize{Examples of theories determined by covariance, field content and symmetry.}}
\label{Table1}
\end{table}
In most cases the solution is unique: the theories listed in the last column of Table 1 are uniquely determined by power counting, field content and symmetry. In all cases, the procedure of finding the most general theory satisfying these very few field theoretical constraints is not only a field theory exercise. All the theories determined in this way are physically relevant. Even topological quantum field theories like the 3D Chern-Simons model are relevant for describing physical phenomena, like quantum Hall systems \cite{Dunne:1998qy,tong}. The list of examples is much longer than the three given in Table 1, and may concern more than one gauge field, like BF theories \cite{Horowitz:1989ng,Birmingham:1991ty,Amoretti:2013xya}, which in $D$ spacetime dimensions involve, besides the gauge field $A_\mu(x)$, a $(D-2)$-form. Like Chern-Simons theory, BF models are important in condensed matter physics. In particular, when a boundary is introduced \cite{Walker:2012mcd,Kapustin:2014gua,Blasi:2019wpq,Amoretti:2014iza,Maggiore:2017vjf,Blasi:2011pf,Amoretti:2012hs}, the main properties of the edge states of Fractional Quantum Hall Effect and of the Topological Insulators are recovered \cite{Cho:2010rk,Hasan:2010xy,Bertolini:2022sao}. 
 Recently, interesting consequences arose from considering the infinitesimal diffeomorphisms, listed in the first row of Table 1, corresponding to a particular choice of the vector parameter $\xi_\mu(x)$ \cite{Bertolini:2022ijb,Blasi:2022mbl,Bertolini:2023sqa,Afxonidis:2023pdq}~:
\be\label{longdiff}
\delta h_{\mu\nu}=\partial_\mu\xi_\nu+\partial_\nu\xi_\mu
\quad\xrightarrow{\xi_\mu\propto\partial_\mu\phi}\quad
\delta h_{\mu\nu}=\partial_\mu\partial_\nu\phi\ ,
\ee
which are also known as longitudinal diffeomorphisms \cite{Dalmazi:2020xou}. The transformation \eqref{longdiff} is the covariant extension of the transformation involving space indices only
\be\label{spacefracton}
\delta h_{ij}=\partial_i\partial_j\phi\ .
\ee
The symmetry transformation \eqref{spacefracton} defines fractons, which are quasiparticles characterized by their limited mobility, when in isolation, which basically means that only fracton dipoles can move \cite{Pretko:2016lgv,Pretko:2016kxt,Prem:2017kxc,Nandkishore:2018sel,Pretko:2020cko}. The covariant generalization \eqref{longdiff} of \eqref{spacefracton} allows to recover all the defining properties of fractons described in \cite{Pretko:2016lgv}, without any additional constraint \cite{Bertolini:2022ijb}. 
In other words, requiring the few quantum field theory ingredients of symmetry, covariance, locality and power counting leads, once again, to a coherent and physically relevant quantum field theory.
To test the goodness of the method, we set ourselves the problem of finding the most general 3D theory
of a generic 9-components tensor field $a_{\mu\nu}(x)$, invariant under the transformation 
\be\label{newsymm}
\delta a_{\mu\nu}=\partial_\mu\xi_\nu\ ,
\ee
aiming to complete Table 1 with the row
\begin{table}[H]
\centering
	\begin{tabular}{|c|c|c|c|}
	\hline
	{\bf dimensions}&{\bf field content}&{\bf symmetry} &{\bf theory} \\\hline
3D&$a_{\mu\nu}$&$\delta a_{\mu\nu}=\partial_\mu\xi_\nu $&?\\\hline
	\end{tabular}
\caption{\footnotesize{Searching for a new 3D theory.}}
\label{Table2}
\end{table}
The problem, which has not been addressed until now, has at least a twofold interest. From a pure field theoretical point of view, the theory we are looking for involves the symmetric part of $a_{\mu\nu}(x)$ and its antisymmetric component which, in 3D, is dual to a vector. It is a gauge field theory which needs a vectorial Nakanishi-Lautrup field \cite{Nakanishi:1966zz,Lautrup:1967zz}, due to the vectorial nature of the gauge parameter $\xi_\mu(x)$ in \eqref{newsymm}. 
Hence, the problem is similar to the gauge fixing of Linearized Gravity \cite{Blasi:2015lrg,Gambuti:2020onb,Gambuti:2021meo,Bertolini:2023juh}, but more complex due to the presence of the antisymmetric component of the gauge tensor field, whose gauge transformation is not the standard one. Hence, the most general gauge fixing condition involves several gauge parameters, corresponding to different possibilities. The problem of finding the most general gauge fixing term which allows to compute the propagators of the theory is non-trivial and, as we shall show, has non-intuitive solutions. Related to this, is the question of the number of degrees of freedom of the theory, which in a gauge field theory is not simply the number of components of the gauge field minus the number of gauge conditions, which for a tensor gauge field $a_{\mu\nu}(x)$ in 3D transforming as \eqref{newsymm} would lead to $3\times 3 - 3 = 6$. We will see that actually the number of degrees of freedom is half this number. 
In addition to this motivation, there is another one, which has a more physical relevance. As we listed in Table 1, the standard gauge theory which can be constructed in 3D with a vector gauge field is the Chern-Simons theory, whose edge states are the chiral modes of quantum Hall systems \cite{tong,Anderson:1972pca,Sachdev:2012dq,Fradkin:2013sab,Bertolini:2021iku}. It is natural to expect that the the 3D theory of a generic rank-2 tensor field would be a higher rank generalization of the Chern-Simons theory, coupled to a vector field in a way determined by the transformation \eqref{newsymm}. The higher rank Chern-Simons theory for a symmetric tensor field has been studied in \cite{Bertolini:2024yur}. There, the generalized Chern-Simons theory was characterized as the most general one invariant under the fractonic transformation \eqref{longdiff}, and an interesting coexistence of fractonic as well as Hall-like behaviours was evidenced. 
In our case the field content is different and the action is defined by a more general symmetry transformation.
It is interesting to determine whether the strange coexistence of Hall and fractonic aspects is preserved or whether it is modified, and if so how. Furthermore, in the context of fracton models, generalized non-symmetric rank-2 tensor fields are sometimes taken into account. For instance, they can emerge from dipole algebra \cite{Afxonidis:2023pdq,Caddeo:2022ibe,Huang:2023zhp}, also applied in the context of superfluids \cite{Glodkowski:2024ova}, or in connection with the so-called fracton-elasticity duality \cite{Pretko:2017kvd,Pretko:2019omh}, as generalizations of the standard case and mainly from a non-covariant point of view \cite{Gromov:2019waa,Grosvenor:2021hkn}. These examples motivate even more the covariant approach adopted in this paper, which might allow to recover unexpected fracton behaviours in non-standard quantum field  theories.
\\

The paper is organized as follows. In Section 2 the model is presented, with the fields, the symmetries and the most general local invariant action. The equations of motion are derived, and are written in terms of an invariant higher rank field strength, which will be important for the definition of generalized electric and magnetic fields. In Section 3 the energy-momentum tensor is computed. Unexpectedly, this turns out to vanish on-shell, which renders this theory very close to a topological quantum field theory \cite{Birmingham:1991ty}, despite the fact that it depends on the spacetime (flat) metric. The most general gauge fixing and the propagators are computed in Section 4. This is a non-trivial task, due to the number of fields involved and to their tensorial nature. We shall see that it is possible to gauge-fix the theory in many ways, some of which are not immediately evident. The counting of the degrees of freedom of the theory is performed in Section 5, where it is shown that they are half of the na\"ive counting. Section 6 contains the physical interpretation of the theory, which is the main motivation for this work. A fractonic behaviour is pointed out, which is surprising, since the defining symmetry of the theory \eqref{newsymm} is not the standard fractonic one \eqref{longdiff}. A generalized electromagnetism and conserved currents are found, from which fractons emerge naturally, in terms of a dipolar structure of this theory. We finally draw our conclusions in Section 7.\\

Our notations and conventions are as follows. We work in flat Minkowskian spacetime, whose metric is $\eta_{\mu\nu}=\mbox{diag}(-1,1,1)$. Greek indices run over spacetime $\mu,\nu,\rho,...=\{0,1,2\}$, while latin indices concern space only 
$i,j,k,...=\{1,2\}$. The Levi-Civita symbol $\epsilon_{\mu\nu\rho}$ is such that $\epsilon_{012}=1$, and the following contractions hold

	\be
		\begin{split}
		\epsilon_{\mu\nu\rho}\epsilon^{\alpha\beta\rho}=-(\delta^\alpha_\mu\delta^\beta_\nu-\delta^\alpha_\nu\delta^\beta_\nu) 
		&\quad;\quad
		\epsilon^{\alpha\beta\mu}\epsilon_{\alpha\beta\nu}=-2\delta^\mu_\nu\\
		\epsilon_{ab0}\epsilon^{cd0}=-(\delta^c_a\delta^d_b-\delta^c_b\delta^d_a) &\quad;\quad
		\epsilon^{ab0}\epsilon_{cb0}=-\delta^a_c\ .
		\end{split}
	\ee

\section{The model\label{sec-non simm}}

\subsection{Fields, symmetries and invariant action}

We consider a generic rank-2 tensor field $a_{\mu\nu}(x)$ which transforms as follows
\be\label{dxi}
	\delta a_{\mu\nu}=\partial_\mu \xi_\nu\ ,
\ee
where $\xi_\mu(x)$ is a local parameter. The tensor $a_{\mu\nu}(x)$ can be decomposed in its symmetric and antisymmetric parts
\be
a_{\mu\nu} = h_{\mu\nu} + \bar h_{\mu\nu}\ ,
\label{adecom}\ee
 where
\be
h_{\mu\nu}=\frac{1}{2}(a_{\mu\nu}+a_{\nu\mu})
\quad ; \quad
\bar h_{\mu\nu}=\frac{1}{2}(a_{\mu\nu}-a_{\nu\mu})\ .
\ee
In 3D the antisymmetric tensor field $\bar h_{\mu\nu}(x)$ is dual to a vector, so that we can write
\be\label{a}
	a_{\mu\nu}=h_{\mu\nu}+\epsilon_{\mu\nu\rho}a^\rho\ .
\ee
On $h_{\mu\nu}(x)$ and $a^\rho(x)$ the transformation \eqref{dxi} reads, after a rescaling of $\xi_\mu(x)$,
	\begin{empheq}{align}
	\delta h_{\mu\nu}&=\partial_{\mu}\xi_\nu+\partial_\nu\xi_\mu\label{diff}\\
	\delta a^\rho&=-\epsilon^{\rho\alpha\beta}\partial_\alpha\xi_\beta\label{da}\ .
	\end{empheq}
Notice that the symmetric tensor field $h_{\mu\nu}(x)$ transforms under infinitesimal diffeomorphisms \eqref{diff}, and that, for a particular choice of $\xi_\mu(x)$, we recover the covariant fracton transformation (a.k.a. longitudinal diffeomorphisms) studied in \cite{Bertolini:2022ijb,Blasi:2022mbl,Bertolini:2023sqa,Afxonidis:2023pdq,Bertolini:2023juh,Bertolini:2023wie,Pena-Benitez:2023aat}
	\be\label{dfract}
	\delta a_{\mu\nu}=\partial_\mu\xi_\nu\quad\xrightarrow{\xi_\mu\propto\partial_\mu\phi}\quad\delta a_{\mu\nu}=\partial_\mu\partial_\nu\phi\quad\Rightarrow\quad \delta h_{\mu\nu}=\partial_\mu\partial_\nu\phi\quad;\quad\delta a^{\rho}=0\ .
	\ee
Nevertheless, we shall never restrict to \eqref{dfract}, but we will always keep the full transformations \eqref{diff} and \eqref{da}, which encodes the infinitesimal diffeomorphisms on the tensor filed $h_{\mu\nu}(x)$.
Assigning unitary canonical mass dimension to the tensor field $a_{\mu\nu}(x)$ and, hence, to its components
\be
	[h_{\mu\nu}]=[a^{\rho}]=1\ ,
\ee
one can find that the most general local action invariant under \eqref{diff} and \eqref{da} and respecting power counting is
\be\label{Sinv}
	S_{inv}=\int d^3x\left(\epsilon^{\mu\nu\rho}h_{\mu}^{\;\lambda}\partial_\nu h_{\rho\lambda}-\epsilon^{\mu\nu\rho}a_\mu\partial_\nu a_\rho+2h_{\mu\nu}\partial^\mu a^\nu-2h\partial_\mu a^\mu\right)\ ,
	\ee
where $h(x)$ is the trace of $h_{\mu\nu}(x)$
\be
	h\equiv\eta^{\mu\nu}h_{\mu\nu}\ .
	\ee	
It is convenient to further decompose $h_{\mu\nu}(x)$ in its traceless and trace part
\be
h_{\mu\nu}=\tilde h_{\mu\nu}+\frac{1}{3}\eta_{\mu\nu}h\ ,
\label{hdecomp}\ee
where
\be
\eta^{\mu\nu}\tilde h_{\mu\nu}=0\ ,
\label{h traceless}\ee
which, from \eqref{diff}, transforms as
\be
\delta \tilde h_{\mu\nu}=\partial_{\mu}\xi_\nu+\partial_\nu\xi_\mu-\frac{2}{3}\eta_{\mu\nu}\partial_\lambda\xi^\lambda\quad;\quad\delta h=2\partial_\lambda\xi^\lambda\ .
\ee
The invariant action $S_{inv}$ \eqref{Sinv} can be written as 
\be\label{Sinv-traceless}
	S_{inv}=\int d^3x\left(\epsilon^{\mu\nu\rho}\tilde h_{\mu}^{\;\lambda}\partial_\nu \tilde h_{\rho\lambda}-\epsilon^{\mu\nu\rho}a_\mu\partial_\nu a_\rho+2\tilde h_{\mu\nu}\partial^\mu a^\nu-\tfrac{4}{3}h\partial_\mu a^\mu\right)\ ,
	\ee
from which the trace $h(x)$ appears to be a Lagrange multiplier, therefore not representing a physical degree of freedom. In fact
	\be\label{diva=0}
	\frac{\delta S_{inv}}{\delta h}=0\quad\Rightarrow\quad \partial_\mu a^\mu=0\ .
	\ee

\subsection{Higher rank field strength}

From the tensor field $a_{\mu\nu}(x)$ it is possible to define a higher rank field strength
\begin{align}
	G_{\mu\nu\rho}=
	&\;\partial_\mu a_{\rho\nu}+\partial_\nu a _{\rho\mu}-\partial_\rho(a_{\mu\nu}+a_{\nu\mu})\nonumber\\
	=&\;F_{\mu\nu\rho}+\epsilon_{\rho\mu\lambda}\partial_\nu a^\lambda+\epsilon_{\rho\nu\lambda}\partial_\mu a^\lambda\label{G}\ ,
	\end{align}
where $F_{\mu\nu\rho}(x)$ depends only on $h_{\mu\nu}(x)$
\be
F_{\mu\nu\rho}= \partial_\mu h_{\nu\rho}+\partial_\nu h_{\mu\rho}
-2\partial_\rho h_{\mu\nu}\ .
\label{F}\ee
Hence, $G_{\mu\nu\rho}(x)$, built in terms of the generic tensor $a_{\mu\nu}(x)$, generalizes the higher rank field strength $F_{\mu\nu\rho}(x)$ in terms of which a Maxwell theory of fractons  can be given \cite{Bertolini:2022ijb,Maggiore:2018bxr}. The field strength \eqref{G} is invariant under \eqref{dxi}
	\be
	\delta G_{\mu\nu\rho}=0\ ,
\label{Ginv}	\ee	
is symmetric under the exchange of the first two indices 
\be
	G_{\mu\nu\rho}=	G_{\nu\mu\rho}\ ,
\label{Gsimm}\ee
and satisfies the cyclicity property
	\be
	G_{\mu\nu\rho}+G_{\nu\rho\mu}+G_{\rho\mu\nu}=0\ .
\label{Gcicl}	\ee
It is useful to define a traceless field strength
\be\label{defGtraceless}
		\begin{split}
		\tilde G_{\mu\nu\rho}&\equiv G_{\mu\nu\rho}-\frac{1}{4}\left(2\eta_{\mu\nu}G^\lambda_{\ \lambda\rho}-\eta_{\mu\rho}G^\lambda_{\ \lambda\nu}-\eta_{\nu\rho}G^\lambda_{\ \lambda\mu}\right)\ ,
		\end{split}
	\ee
which, besides being invariant \eqref{Ginv}, symmetric \eqref{Gsimm} and cyclic \eqref{Gcicl}, it is completely traceless, in the sense that
\be
\eta_{\mu\nu}\tilde G^{\mu\nu\rho}=\eta_{\mu\rho}\tilde G^{\mu\nu\rho}=\eta_{\nu\rho}\tilde G^{\mu\nu\rho}=0\ .\label{Gtraceless}
\ee

\subsection{Equations of motion}

From the invariant action $S_{inv}$ \eqref{Sinv} we get the equations of motion (EoM) 
\bea
\frac{\delta S_{inv}}{\delta a_\alpha}&=&
-2\epsilon^{\alpha\mu\nu}\partial_\mu a_\nu+2\partial^\alpha h-2\partial_\mu h^{\mu\alpha}
=-G_\mu^{\ \mu\alpha}
\label{eom a}\\
\frac{\delta S_{inv}}{\delta h_{\alpha\beta}} 
&=&
\epsilon^{\alpha\mu\nu}\partial_\mu h_\nu^{\;\beta}+\epsilon^{\beta\mu\nu}\partial_\mu h_\nu^{\;\alpha}+\partial^\alpha a^\beta+\partial^\beta a^\alpha-2\eta^{\alpha\beta}\partial_\mu a^\mu\nonumber \\
&=&
\frac{1}{3}\left(\epsilon^{\alpha\mu\nu}\tilde G^\beta_{\ \mu\nu}+\epsilon^{\beta\mu\nu}\tilde G^\alpha_{\ \mu\nu}\right)-\frac{4}{3}\eta^{\alpha\beta}\partial_\lambda a^{\lambda}\label{eom h}\ ,
\eea
from which we derive the EoM for the traceless tensor field $\tilde h_{\alpha\beta}(x)$ and for the trace $h(x)$
	\begin{align}
	\frac{\delta S_{inv}}{\delta \tilde h_{\alpha\beta}}&=\frac{1}{3}{\left(\epsilon^{\alpha\mu\nu}\tilde G^\beta_{\ \mu\nu}+\epsilon^{\beta\mu\nu}\tilde G^\alpha_{\ \mu\nu}\right)}\label{eomh}\\
	\frac{\delta S_{inv}}{\delta h}&=-4\partial_\mu a^\mu\ .\label{eomTrh}
	\end{align}
Making explicit space and time indices and going on-shell, from \eqref{eomh} we get
\bi
	\item $\pmb{\alpha=\beta=0}$
		\be
		\epsilon^{0mn}\tilde G^0_{\ mn}=0\quad\Rightarrow\quad\tilde G_{0mn}=\tilde G_{0nm}\ ,
		\ee
and hence		
		\be\label{vincolo h00}
	\tilde G_{0mn}=-\frac{1}{2}\tilde G_{mn0}\ ,
		\ee
due to the cyclicity property \eqref{Gcicl}.
	\item  $\pmb{\alpha=a,\ \beta=0}$
		\be
\epsilon^{0mn}\tilde G^a_{\ mn}+\frac{3}{2}\epsilon^{0am}\tilde G^n_{\ nm}=0\ ,
		\ee
where the cyclicity \eqref{Gcicl} and the tracelessness \eqref{Gtraceless} properties have been used.
	\item  $\pmb{\alpha=a,\ \beta=b}$
		\be
\epsilon^{0am}\tilde G^b_{\ m0}+\epsilon^{0bm}\tilde G^a_{\ m0}=0\ ,
		\ee
after using \eqref{vincolo h00}.
	\ei
For what concerns the EoM  \eqref{eom a} for the vector field $a_\alpha(x)$, on-shell we have
	\bi
	\item $\pmb{\alpha=0}$
		\be
	G_m^{\ m0} = 0\ ,
		\ee

	\item $\pmb{\alpha=a}$
		\be
		G_0^{\ 0 a}+G_m^{\ m a}= 0\ .
		\ee
	\ei

\section{Energy-momentum tensor}

In 3D, the theory of a generic rank-2 tensor $a_{\mu\nu}(x)$ invariant under the transformation \eqref{dxi} has the rather peculiar property of being ``almost'' topological, in a sense that we are going to clarify. Looking at the action $S_{inv}$ \eqref{Sinv}, we see a non-trivial dependence on the spacetime metric and one would hardly claim that this theory has anything to do with a topological quantum field theory, whose primary feature is its independence on the spacetime metric. In fact, the 3D prototype of a topological quantum field theory is Chern-Simons theory, which, in the abelian case, is
\be
S_{CS}=\int d^3x\; \epsilon^{\mu\nu\rho}A_\mu\partial_\nu A_\rho\ ,
\label{cs}\ee
and $A_\mu(x)$ is the standard vector gauge field. The action \eqref{cs} does not depend on the spacetime metric $g_{\mu\nu}(x)$, with the important consequence that the observables of the theory are global, and not local, like for instance the Donaldson polynomials invariants and the Floer groups of the 3D manifolds \cite{ Witten:1988ze,Witten:1988hf}. Actually, the action $S_{inv}$ is quite similar to the Chern-Simons one \eqref{cs}. In fact, keeping the original tensor field $a_{\mu\nu}(x)$ without making the decomposition \eqref{adecom} or, equivalently \eqref{a}, the most general action invariant under \eqref{dxi} is
\be
S_{inv} = \int d^3x\; \epsilon^{\mu\nu\rho}\;a_\mu^{\ \lambda}\partial_\nu a_{\rho\lambda}\ ,
\label{Sinva}\ee
which is the same $S_{inv}$ \eqref{Sinv} once the decomposition \eqref{a} is done. The invariant action in the form \eqref{Sinva} closely reminds the Chern-Simons action \eqref{cs}, with the difference that $S_{inv}$ has a linear metric dependence
\be
S_{inv} = \int d^3x\; \epsilon^{\mu\nu\rho}g^{\alpha\beta}\;a_{\mu\alpha}\partial_\nu a_{\rho\beta}\ ,
\label{Sinvag}\ee
which is mild, compared, for instance, with the cubic metric dependence of the Maxwell theory
\be
S_{Max}=-\frac{1}{4}\int d^4x\sqrt{-g}\;g^{\mu\nu}g^{\rho\sigma}F_{\mu\rho}F_{\nu\sigma}\ .
\label{maxconmetrica}\ee
The similarity between $S_{inv}$ and $S_{CS}$ is not only formal. The energy-momentum tensor 
 \be
 T_{\mu\nu}= -\frac{2}{\sqrt{-g}}\frac{\delta S_{inv}}{\delta g^{\mu\nu}}
 \label{Tmunu}\ee
 for the Chern-Simons theory, which does not depend on the metric, evidently vanishes
 \be
  T_{\mu\nu}^{CS}=0\ .
 \label{Tmunucs}\ee
 For the theory we are here considering, it is convenient to switch to momentum space\footnote{$\hat\Phi (p)$ is the Fourier transform of $\Phi(x)$.}, where the action reads
\be
S_{inv} = -i\int d^3p\; \epsilon^{\mu\nu\rho}\hat a_\mu^{\ \lambda}(p)p_\nu\hat a_{\rho\lambda}(-p)\ .
\label{SinvF}\ee
The Fourier transform of the energy-momentum tensor therefore is
\be
\hat T_{\mu\nu}(p) = i \epsilon^{\alpha\beta\gamma}\left[
\hat a_{\alpha\mu}(p)p_\beta\hat a_{\gamma\nu}(-p) + 
\hat a_{\alpha\nu}(p)p_\beta\hat a_{\gamma\mu}(-p)
\right]\ ,
\label{Tmunua}\ee
which, using the EoM
\be
\frac{\delta S_{inv}}{\delta\hat a_{\mu\nu}(-p)} = -2i\epsilon^{\mu\alpha\beta}p_\beta\hat a_\alpha^{\ \nu}(p)
\label{eoma}\ee
reads
\be
\hat T_{\mu\nu}(p) = -\frac{1}{2}\left[
\hat a_{\lambda\mu}(-p) \frac{\delta S_{inv}}{\delta\hat a_\lambda^{\ \nu}(-p)} +
\hat a_{\lambda\nu}(-p) \frac{\delta S_{inv}}{\delta\hat a_\lambda^{\ \mu}(-p)}
\right]\ ,
\label{Tmunuonsgell}\ee
which vanishes, once the EoM are used, that is on-shell
\be
\frac{\delta S_{inv}}{\delta\hat a_{\mu\nu}(p)} = 0\ .
\label{onshell}\ee 
Thus, the mild dependence of $S_{inv}$ \eqref{Sinva} on the metric reflects into the fact that the energy-momentum tensor vanishes on-shell, which renders this theory ``almost'' topological, as we said. It is also readily seen that the energy-momentum tensor \eqref{Tmunua} is, on-shell, a conserved current
\be
p_\nu \hat T^{\mu\nu} =0\ ,
\label{Tconserved}\ee
as it should.

\section{Gauge fixing and propagators}

The model we are considering in this paper is defined by the transformation \eqref{dxi}, which encodes the infinitesimal diffeomorphisms \eqref{diff}. Therefore, it is a gauge field theory which needs to be gauge fixed in order to be well defined. In particular, the gauge fixing allows the existence of the Green functions generating functionals. In this Section we perform the gauge fixing procedure, which reveals some peculiar aspects, and we compute the 2-points Green function, $i.e.$ the propagators of the theory. We then discuss the pole structure of the progataros, which is a powerfiul tool to get insights of this theory. The transformation \eqref{dxi}, or equivalently \eqref{diff} and \eqref{da}, depends on the vector local gauge parameter $\xi^\mu(x)$. As a consequence, the most general gauge fixing is
	\be
\kappa_1\partial^\nu h_{\mu\nu}+\kappa_2\partial_\mu h+\kappa_3\epsilon_{\mu\nu\rho}\partial^\nu a^\rho=0\ ,
	\label{gfcond}\ee
which is implemented by a Lagrange vector field $b^\mu(x)$ \cite{Nakanishi:1966zz,Lautrup:1967zz}
	\be\label{Sgf}
	S_{gf}=\int d^3x\,b^\mu\left[
\kappa_1\partial^\nu h_{\mu\nu}+\kappa_2\partial_\mu h+\kappa_3\epsilon_{\mu\nu\rho}\partial^\nu a^\rho
	+\frac{\kappa}{2}b_\mu
	\right]\ ,
	\ee
where $\{\kappa,\kappa_1,\kappa_2,\kappa_3\}$ are gauge parameters. However, since power counting implies that $b^\mu(x)$ has canononical mass dimension  $[b]=1$, we have that the gauge parameter $\kappa$ is not dimensionless $[\kappa]=1$. This would lead to unavoidable infrared divergences \cite{Birmingham:1991ty,Alvarez-Gaume:1989ldl}, exactly as it happens in Chern-Simons theory, which is for this reason defined only in the Landau gauge, which therefore we shall adopt~:
\be
	\kappa=0\ .
	\ee
The gauge fixed action
\be
S_{tot} = S_{inv} + S_{gf}\ ,
\label{Stot}\ee
where $S_{inv}$ is the invariant action \eqref{Sinv}, in momentum space reads
\be
		\begin{split}
		 S_{tot}=\int d^3p&\left(\hat h_{\mu\nu},\hat a_{\mu},\hat b_{\mu}\right)\left(
			\begin{array}{ccc}
			-\frac{1}{4}\mathcal{E}^{\mu\nu,\alpha\beta}&-\frac{1}{2}n^{\mu\nu,\alpha}&\frac{1}{4}n_{_{(\kappa_1,\kappa_2)}}^{\mu\nu,\alpha}\\
			-\frac{1}{2}n^{*\alpha\beta,\mu}&i\epsilon^{\mu\lambda\alpha}p_\lambda&-\frac{i}{2}\kappa_3\epsilon^{\mu\lambda\alpha}p_\lambda\\
			\frac{1}{4}n_{_{(\kappa_1,\kappa_2)}}^{*\alpha\beta,\mu}&-\frac{i}{2}\kappa_3\epsilon^{\mu\lambda\alpha}p_\lambda&0\\
			\end{array}\right)
		\left(
			\begin{array}{c}
			\hat h_{\alpha\beta}\\
			\hat a_{\alpha}\\
			\hat b_{\alpha}
			\end{array}\right)\\
		\equiv\int d^3p\,&\hat\Phi_A(p) M^{AB}(p)\hat\Phi_B(-p)\ ,
		\end{split}
	\ee
where
	\begin{empheq}{align}
	\mathcal{E}^{\mu\nu,\alpha\beta}&\equiv ip_\lambda\left(\epsilon^{\mu\lambda\alpha}\eta^{\nu\beta}+\epsilon^{\mu\lambda\beta}\eta^{\nu\alpha}+\epsilon^{\nu\lambda\alpha}\eta^{\mu\beta}+\epsilon^{\nu\lambda\beta}\eta^{\mu\alpha}\right)\\
	n^{\mu\nu,\alpha}&\equiv i\left(p^\mu\eta^{\nu\alpha}+p^\nu\eta^{\mu\alpha}-2p^{\alpha}\eta^{\mu\nu}\right)\\
	n^{\mu\nu,\alpha}_{_{(\kappa_1,\kappa_2)}}&\equiv i\left[\kappa_1\left(p^\mu\eta^{\nu\alpha}+p^\nu\eta^{\mu\alpha}\right)+2\kappa_2p^{\alpha}\eta^{\mu\nu}\right]\ ,
	\end{empheq}
notice that $n^{\mu\nu,\alpha}_{_{(\kappa_1,-\kappa_1)}}=\kappa_1 n^{\mu\nu,\alpha}$. The momentum space matrix propagator $D_{AB}(p)$ solves
\be\label{prop eq}
	M^{AC}D_{CB}=I^A_{\ B}\ ,
	\ee
where 
	\be
	D_{BC}\equiv\left(
		\begin{array}{ccc}
		{H}_{\alpha\beta,\rho\sigma}&{H}_{\alpha\beta,\rho}&{\mathcal{H}}_{\alpha\beta,\rho}\\
		{H}^{*}_{\rho\sigma,\alpha}&A_{\alpha\rho}&\mathcal{A}_{\alpha\rho}\\
		{\mathcal{H}}^*_{\rho\sigma,\alpha}&\mathcal{A}^*_{\rho\alpha}&B_{\alpha\rho}\\
		\end{array}\right)\ ,
	\ee
		\begin{align}
		&{H}_{\alpha\beta,\rho\sigma}(p)\equiv \langle \hat h_{\alpha\beta}(p)\hat h_{\rho\sigma}(-p)\rangle\quad;\quad
		{H}_{\alpha\beta,\rho}(p)\equiv \langle \hat h_{\alpha\beta}(p)\hat a_{\rho}(-p)\rangle\quad;\quad
		{\mathcal{H}}_{\alpha\beta,\rho}(p)\equiv\langle \hat h_{\alpha\beta}(p)\hat b_{\rho}(-p)\rangle\nonumber\\
		&A_{\alpha\rho}(p)\equiv\langle \hat a_{\alpha}(p)\hat a_{\rho}(-p)\rangle\quad;\quad
		\mathcal{A}_{\alpha\rho}(p)=\mathcal{A}^*_{\rho\alpha}\equiv\langle \hat b_{\alpha}(p)\hat a_{\rho}(-p)\rangle\quad;\quad
		B_{\alpha\rho}(p)\equiv\langle \hat b_{\alpha}(p)\hat b_{\rho}(-p)\rangle\label{propagatori}
		\end{align}
and $I^A_{\ B}$ is the identity matrix
\bea
I^{\mu\nu}_{\ \ \alpha\beta} &=& \frac{1}{2}
(\delta^\mu_\alpha\delta^\nu_\beta + \delta^\mu_\beta\delta^\nu_\alpha)\\
I^{\mu}_{\ \alpha}  &=& \delta^\mu_\alpha\ .
\eea
The matrix equation \eqref{prop eq} is solved in Appendix \ref{App-Prop} and we get the following propagators
	\begin{empheq}{align}
	{H}_{\alpha\beta,\rho\sigma}&=\frac{1}{4p^2}\tilde{\mathcal{E}}_{\alpha\beta,\rho\sigma}+\frac{\kappa_3^2}{(\kappa_1-\kappa_3)^2p^4}\bar{\mathcal{E}}_{\alpha\beta,\rho\sigma}\label{Delta'}\\
	{H}_{\alpha\beta,\rho}&=\frac{i}{2p^2}t_{\alpha\beta}\,p_{\rho}+\frac{i\kappa_1 \kappa_3}{(\kappa_1-\kappa_3)^2}\frac{1}{p^2}\left(t_{\alpha\rho}\,p_\beta+t_{\beta\rho}\,p_{\alpha}\right)-\frac{i\kappa_2}{\kappa_1+\kappa_2}\frac{p_\alpha p_\beta}{p^4} p_{\rho}\label{D'}\\
	{\mathcal{H}}_{\alpha\beta,\rho}&=\frac{2i}{(\kappa_1+\kappa_2)}\frac{p_\alpha p_\beta }{p^4}p_{\rho}+\frac{2i}{(\kappa_1-\kappa_3)p^2}\left(t_{\alpha\rho}\,p_\beta+t_{\beta\rho}\,p_\alpha\right)\label{d'}\\
	A_{\alpha\rho}&= -\frac{i\kappa_1^2}{(\kappa_1-\kappa_3)^2}\frac{p^\lambda\epsilon_{\alpha\lambda\rho}}{p^2}\label{A'}\\
	\mathcal{A}_{\alpha\rho}&=-\frac{2i}{(\kappa_1-\kappa_3)}\frac{p^\lambda\epsilon_{\alpha\lambda\rho}}{p^2}\label{McA'}\\
	B_{\alpha\rho}&= 0\ ,\label{B'}
	\end{empheq}
	where
\begin{empheq}{align}
	t_{\alpha\beta}&\equiv\eta_{\alpha\beta}-\tfrac{p_\alpha p_\beta}{p^2}\\
	\bar{\mathcal{E}}_{\alpha\beta,\rho\sigma}&\equiv ip^\lambda\left(\epsilon_{\alpha\lambda\rho}p_\sigma p_\beta+\epsilon_{\alpha\lambda\sigma}p_\beta p_\rho+\epsilon_{\beta\lambda\sigma}p_\alpha p_\rho+\epsilon_{\beta\lambda\rho}p_\alpha p_\sigma\right)\\
	\tilde{\mathcal{E}}_{\alpha\beta,\rho\sigma}&\equiv\mathcal{E}_{\alpha\beta,\rho\sigma}-\frac{\bar{\mathcal{E}}_{\alpha\beta,\rho\sigma}}{p^2}= ip^\lambda\left(\epsilon_{\alpha\lambda\rho}t_{\beta\sigma}+\epsilon_{\alpha\lambda\sigma}t_{\beta\rho}+\epsilon_{\beta\lambda\sigma}t_{\alpha\rho}+\epsilon_{\beta\lambda\rho}t_{\alpha\sigma}\right)\ .
\end{empheq}
We observe poles at
\be\label{poli}
	\kappa_2=-\kappa_1\quad;\quad \kappa_3=\kappa_1\ ,
\ee
which identify the gauge parameters in \eqref{gfcond} not corresponding to a good gauge fixing. All other combinations of $\kappa_1$, $\kappa_2$ and $\kappa_3$ gauge fix the theory defined by the action $S_{inv}$ \eqref{Sinv}. It is interesting to notice that some cases are not readily recognized as ``good'' gauge fixing. For instance, interesting situations are
\begin{itemize}
\item 
$\pmb{\kappa_1=0}$, which, from \eqref{poli}, implies that $\kappa_2\neq0$ and $\kappa_3\neq0$. The corresponding gauge fixing term $S_{gf}$ is 
	\be\label{gf0kl}
	S_{gf}(0,\kappa_2,\kappa_3)=\int d^3x\,b^\mu\left[\kappa_2\partial_\mu h+\kappa_3\epsilon_{\mu\nu\rho}\partial^\nu a^\rho\right]\ ,
	\ee
and corresponding propagators are
	\begin{empheq}{align}
	{H}_{\alpha\beta,\rho\sigma}&=\frac{1}{4p^2}\tilde{\mathcal{E}}_{\alpha\beta,\rho\sigma}+\frac{1}{p^4}\bar{\mathcal{E}}_{\alpha\beta,\rho\sigma}\label{Delta'0kl}\\
	{H}_{\alpha\beta,\rho}&=\frac{i}{2p^2}t_{\alpha\beta}\,p_{\rho}-i\frac{p_\alpha p_\beta}{p^4} p_{\rho}\label{D'0kl}\\
	{\mathcal{H}}_{\alpha\beta,\rho}&=\frac{2i}{\kappa_2}\frac{p_\alpha p_\beta}{p^4} p_{\rho}-\frac{2i}{\kappa_3p^2}\left(t_{\alpha\rho}\,p_\beta+t_{\beta\rho}\,p_\alpha\right)\label{d'0kl}\\
	A_{\alpha\rho}&=0\label{A'0kl}\\
	\mathcal{A}_{\alpha\rho}&=\frac{2i}{\kappa_3}\frac{p^\lambda\epsilon_{\alpha\lambda\rho}}{p^2}\label{McA'0kl}\\
	B_{\alpha\rho}&= 0\ .\label{B'0kl}
	\end{empheq}
\item 
$\pmb{\kappa_2=\kappa_3=0}$, which, from \eqref{poli}, implies $\kappa_1\neq0$. Hence, in this case the gauge fixing term involves only $h_{\mu\nu}(x)$. For instance, if $\kappa_1=1$ the gauge fixing term reads
	\be\label{Sgf00}
	S_{gf}(1,0,0)=\int d^3x\,b^\mu\partial^\nu h_{\mu\nu}\ ,
	\ee
and the propagators greatly simplify	
\begin{empheq}{align}
	{H}_{\alpha\beta,\rho\sigma}&=\frac{1}{4p^2}\tilde{\mathcal{E}}_{\alpha\beta,\rho\sigma}
	\label{Delta'00}\\
	{H}_{\alpha\beta,\rho}&=
\frac{i}{2p^2}t_{\alpha\beta}\,p_{\rho}\label{D'00}\\
	{\mathcal{H}}_{\alpha\beta,\rho}&= 
	 \frac{2i}{p^2}\left(t_{\alpha\rho}\,p_\beta+t_{\beta\rho}\,p_\alpha+\frac{p_\alpha p_\beta }{p^2}p_{\rho}\right)
\label{d'00}\\
	A_{\alpha\rho}&= -\frac{i}{p^2}p^\lambda\epsilon_{\alpha\lambda\rho}\label{A'00}\\
	\mathcal{A}_{\alpha\rho}&=-\frac{2i}{p^2}p^\lambda\epsilon_{\alpha\lambda\rho}\label{McA'00}\\
	B_{\alpha\rho}&= 0\ .\label{B'00}
	\end{empheq}
\item 
$\pmb{\kappa_1=1,\ \kappa_2=-\frac{1}{3},\ \kappa_3=0}$. This corresponds to gauge fixing only the traceless component $\tilde h_{\mu\nu}(x)$, which is not obvious to be sufficient to get the propagators. The gauge fixing term is 
	\be\label{Sgf01/3}
	S_{gf}(1,-\tfrac{1}{3},0)=\int d^3x\,b^\mu\partial^\nu\tilde h_{\mu\nu}\ ,
	\ee
and the corresponding propagators are
	\begin{empheq}{align}
	{H}_{\alpha\beta,\rho\sigma}&=
	\frac{1}{4p^2}\tilde{\mathcal{E}}_{\alpha\beta,\rho\sigma}\label{Delta'0tr}\\
	{H}_{\alpha\beta,\rho}&= \frac{i}{2p^2}\eta_{\alpha\beta}\,p_{\rho}\label{D'0tr}\\
	{\mathcal{H}}_{\alpha\beta,\rho}&=
\frac{2i}{p^2}\left(t_{\alpha\rho}\,p_\beta+t_{\beta\rho}\,p_\alpha+\frac{3}{2}\frac{p_\alpha p_\beta }{p^2}p_{\rho}\right)
\label{d'0tr}\\
	A_{\alpha\rho}&= -\frac{i}{p^2}p^\lambda\epsilon_{\alpha\lambda\rho}\label{A'0tr}\\
	\mathcal{A}_{\alpha\rho}&=-\frac{2i}{p^2}p^\lambda\epsilon_{\alpha\lambda\rho}\label{McA'0tr}\\
	B_{\alpha\rho}&= 0\ .\label{B'0tr}
	\end{empheq}
\end{itemize}

\section{Degrees of freedom}

In this Section we shall show that the number of degrees of freedom (DoF) of the theory does not depend on the gauge parameters $\kappa_1$, $\kappa_2$ and $\kappa_3$, as it should, which allows us to fix any value of them, provided that they lie outside the poles \eqref{poli}.
From the invariant action $S_{inv}$ \eqref{Sinv} and the gauge fixing term $S_{gf}$ \eqref{Sgf} we get the following 
momentum space EoM 
\begin{align}
	\frac{\delta S_{tot}}{\delta \hat b_\alpha}&=\kappa_1p_\mu \hat h^{\mu\alpha}+\kappa_2p^\alpha\hat h+\kappa_3\epsilon^{\alpha\mu\nu}p_\mu \hat a_\nu 
	\label{peomb-gf}\\
	\frac{\delta S_{tot}}{\delta \hat a_\alpha}&=-2\epsilon^{\alpha\mu\nu}p_\mu \hat a_\nu+2p^\alpha \hat h-2p_\mu \hat h^{\mu\alpha}+\kappa_3\epsilon^{\alpha\mu\nu}p_\mu \hat b_\nu 
	\label{peoma-gf}\\
\frac{\delta S_{tot}}{\delta \hat h_{\alpha\beta}}&=\epsilon^{\alpha\mu\nu}p_\mu \hat h_\nu^{\,\beta}+\epsilon^{\beta\mu\nu}p_\mu \hat h_\nu^{\,\alpha}+p^\alpha\hat a^\beta+p^\beta\hat a^\alpha-2\eta^{\alpha\beta}p_\mu\hat a^\mu-\kappa_2\eta^{\alpha\beta}p_\mu\hat b^\mu-\frac{\kappa_1}{2}\left(p^\alpha\hat b^\beta+p^\beta \hat b^\alpha\right)\ ,\label{peomh-gf}
	\end{align}
from which, going on-shell, we have
	\begin{align}
	p_\alpha\frac{\delta S_{tot}}{\delta\hat a_\alpha}&=-p^2\hat h+p_\mu p_\nu\hat h^{\mu\nu}=0\\
	p_\alpha\frac{\delta S_{tot}}{\delta\hat b_\alpha}&=\kappa_1 p_\mu p_\nu\hat h^{\mu\nu}+\kappa_2p^2\hat h=0\ ,
	\end{align}
and, for $\kappa_2\neq-\kappa_1$, that is outside the pole \eqref{poli}, we get 
	\begin{align}
	p_\mu p_\nu\hat h^{\mu\nu}&=0\label{pph=0-gf}\\
	p^2\hat h&=0\ .\label{p2h=0-gf}
	\end{align}
Multiplying the $\hat b_\alpha(p)$ EoM \eqref{peomb-gf} by $p^2$, going on-shell and using \eqref{p2h=0-gf}, we find
\be
	\kappa_1p^2p_\mu \hat h^{\mu\alpha} + p^2\kappa_3 \epsilon^{\alpha\mu\nu}p_\mu\hat a_\nu=0\ .
\ee
For $\kappa_3=0$ (hence $\kappa_1\neq0$), we get 
\be
	p^2p_\mu \hat h^{\mu\alpha}=0\ .
\ee
If instead $\kappa_3\neq0$, we have
\be\label{rota=h-gf}
	p^2 \epsilon^{\alpha\mu\nu}p_\mu\hat a_\nu=-\frac{\kappa_1}{\kappa_3} p^2p_\mu\hat h^{\mu\alpha}\ .
\ee
Acting on the $\hat a_\alpha(p)$ EoM \eqref{peoma-gf} with $p^2$ and using \eqref{p2h=0-gf}, on-shell we have
	\be
	-p^2\epsilon^{\alpha\mu\nu}p_\mu\hat a_\nu-p_\mu p^2\hat  h^{\mu\alpha}+\frac{\kappa_3}{2}p^2\epsilon^{\alpha\mu\nu}p_\mu\hat  b_\nu\\
=
\frac{\kappa_1}{\kappa_3}p^2p_\nu\hat  h^{\alpha\nu}-p^2p_\nu\hat  h^{\alpha\nu}+\frac{\kappa_3}{2}p^2\epsilon^{\alpha\mu\nu}p_\mu \hat b_\nu=0\ ,
	\ee
from which	
\be\label{rotb=h-gf}
	p^2\epsilon^{\alpha\mu\nu}p_\mu\hat  b_\nu=2\frac{\kappa_3-\kappa_1}{\kappa_3^2}p^2p_\mu \hat h^{\alpha\mu }\ .
\ee	
Finally, considering
\be\label{ppph=0-gf}
		\begin{split}
		0&=\epsilon_{\beta\gamma\sigma}p^\gamma p_\alpha\frac{\delta S_{tot}}{\delta\hat  h_{\alpha\beta}}\\
		&=\epsilon_{\beta\gamma\sigma}\epsilon^{\beta\mu\nu}p^\gamma p_\alpha p_\mu\hat   h_\nu^{\,\alpha}+p^2\epsilon_{\beta\gamma\sigma}p^\gamma \hat a^\beta-\frac{\kappa_1}{2}p^2\epsilon_{\beta\gamma\sigma}p^\gamma\hat  b^\beta\\
		&=(\delta^\mu_\sigma\delta^\nu_\gamma-\delta^\mu_\gamma\delta^\nu_\sigma)p^\gamma p^\alpha p_\mu  \hat h_{\nu\alpha}-p^2\epsilon_{\sigma\nu\rho}p^\nu\left(\hat a^\rho-\frac{\kappa_1}{2}\hat b^\rho\right)\\
		&=-p^2p^\alpha\hat   h_{\alpha\sigma}+p^2\left(\frac{\kappa_1}{\kappa_3}+\kappa_1\frac{\kappa_3-\kappa_1}{\kappa_3^2}\right)p^\nu \hat  h_{\sigma\nu}\\
		&=\frac{(\kappa_3-\kappa_1)^2}{\kappa_3^2}p^2p^\nu\hat   h_{\nu\sigma}\ ,
		\end{split}
	\ee
where  \eqref{pph=0-gf}, \eqref{rota=h-gf} and \eqref{rotb=h-gf} have been used. Since $\kappa_3\neq\kappa_1$ to avoid the pole \eqref{poli}, from \eqref{ppph=0-gf}, \eqref{rota=h-gf} and \eqref{rotb=h-gf} we have
	\begin{align}
	p^2p^\nu  \hat h_{\nu\sigma}&=0\label{pmuh}\\
	p^2\epsilon^{\mu\nu\rho}p_\nu \hat a_\rho&=0\label{p2dofa}\\
	p^2\epsilon^{\mu\nu\rho}p_\nu \hat b_\rho&=0\ .\label{p2dofb}
	\end{align}
Following the same method adopted in the counting of the DoF of linearized gravity \cite{Blasi:2022mbl,Blasi:2015lrg,Gambuti:2020onb,Gambuti:2021meo,Bertolini:2023juh,Blasi:2017pkk},
 from \eqref{pmuh}, \eqref{p2dofa} and \eqref{p2dofb} we get  
 \begin{align}
	p^\nu\hat h_{\nu\sigma}&=0\label{dofh}\\
	\epsilon^{\mu\nu\rho}p_\nu\hat a_\rho&=0\label{dofa}\\
	\epsilon^{\mu\nu\rho}p_\nu \hat b_\rho&=0\ .
	\end{align}
We can conclude that the DoF of the theory of the rank-2 tensor field $a_{\mu\nu}(x)$ invariant under the transformation \eqref{dxi} are three. In fact, the $3\times3$ components of the tensor field $a_{\mu\nu}(x)$ can be decomposed in a symmetric tensor field $h_{\mu\nu}(x)$ (six components), and a vector field  $a_\mu(x)$ for its antisymmetric part (three components). From \eqref{dofa}, the vector field $a_{\mu}(x)$ can be written in terms of a scalar field $\partial_\mu\varphi(x)$, hence it counts as one DoF. Because of the three gauge conditions \eqref{dofh}, $h_{\mu\nu}(x)$ contributes with three DoF. Finally, we see from the action \eqref{Sinv-traceless} that $h(x)$ is a Lagrange multiplier and therefore it does not count as a DoF. Hence, the number of DoF is three.

\section{Fractonic interpretation}

\subsection{Generalized electromagnetism and currents}

Following the same procedure which in 3D leads to the definition of an electric and magnetic field for Chern-Simons theory \cite{Dunne:1998qy, tong}, from the higher rank theory described by the invariant action \eqref{Sinv}  we can define an ``electric'' tensor field $E^{ab}(x)$ and a vector ``magnetic'' field $B^a(x)$ as follows
\begin{align}
	\frac{\delta S_{inv}}{\delta h_{ab}}&\equiv \frac{1}{2}\left(\epsilon^{0am} E^b_{\ m}+\epsilon^{0bm} E^a_{\ m}\right)\label{eomab=E}\\
	\frac{\delta S_{inv}}{\delta h_{a0}}&\equiv B^a\ ,\label{eoma0=B}
	\end{align}
which generalize the ordinary electric and magnetic fields described, in 3D, by a vector and a scalar field, respectively. This is possible provided that
\begin{align}
	E_{ab}&=\tilde G_{ab0}\label{tensorE}\\
    	B^a&=\frac{2}{3}\epsilon^{0mn}\tilde G^a_{\ mn}
	\label{vectorB}\ ,
	\end{align}
where the EoM \eqref{diva=0} and the solution \eqref{vincolo h00} of the EoM \eqref{eomh} have been used. It is therefore important to remark that the generalized electric and magnetic fields \eqref{tensorE} and \eqref{vectorB} are defined on-shell. From \eqref{tensorE} we see that the generalized electric field is traceless
\be
E^{ab}=\tilde E^{ab}\quad;\quad \eta_{ab}\tilde E^{ab}=0\ .
\label{tracelessE}\ee
Notice indeed that the trace of the field strength $G_\mu^{\ \mu\lambda}(x)$ only intervenes in the EoM of the antisymmetric (vectorial) contribution \eqref{eom a}.
We now couple the theory to a matter current $j^{\mu\nu}(x)$
	\be
	S_J=-\int d^3x j^{\mu\nu}a_{\mu\nu}=-\int d^3x\left(J^{\mu\nu}h_{\mu\nu}+J^\mu a_\mu\right)\ ,
\label{SJ}	\ee
which, as done for the rank-2 tensor $a_{\mu\nu}$ \eqref{a}, is decomposed into its symmetric and antisymmetric components
	\be\label{j}
	j^{\mu\nu}=J^{\mu\nu}-\frac{1}{2}\epsilon^{\mu\nu\rho}J_\rho\quad;\quad J^{\mu\nu}=J^{\nu\mu}\ .
	\ee
The introduction of the matter contribution $S_J$  \eqref{SJ} modifies
the EoM \eqref{eom h} and \eqref{eom a} as follows
	\begin{align}
	J^{\alpha\beta}&=\frac{\delta S_{inv}}{\delta h_{\alpha\beta}}\label{J=eomh}\\
	J^{\alpha}&=\frac{\delta S_{inv}}{\delta a_{\alpha}}\ .\label{J=eoma}
	\end{align}
Since the  electric and magnetic tensor fields \eqref{tensorE} and \eqref{vectorB} are defined on-shell, in particular using the solution \eqref{vincolo h00} of the EoM \eqref{eomh}, we should consistently set 
	\be\label{J00=0}
	J^{00}=\frac{\delta S_{inv}}{\delta h_{00}}=0\ ,
	\ee
so that
	\begin{align}
	J^{ab}=\frac{\delta S_{inv}}{\delta h_{ab}}&\equiv \frac{1}{2}\left(\epsilon^{0am}\tilde E^b_{\ m}+\epsilon^{0bm}\tilde E^a_{\ m}\right)\label{Jab=E}\\
	J^{0a}=\frac{\delta S_{inv}}{\delta h_{a0}}&\equiv B^a\ .\label{Ja0=B}
	\end{align}
Concerning the EoM \eqref{J=eoma} of the vector $a_\mu(x)$, we have
	\begin{align}
	J^0=\frac{\delta S_{inv}}{\delta a_0}&=-G_m^{\ m0}
\label{J0=B}\\
	J^a=\frac{\delta S_{inv}}{\delta a_a}&=-G_\mu^{\ \mu a}\ .\label{Ja=E}
	\end{align}
Expliciting the EoM \eqref{eom h}, from \eqref{J=eomh} we get 
	\be
	J^{\alpha\beta}=\epsilon^{\alpha\mu\nu}\partial_\mu h_\nu^{\;\beta}+\epsilon^{\beta\mu\nu}\partial_\mu h_\nu^{\;\alpha}+\partial^\alpha a^\beta+\partial^\beta a^\alpha-2\eta^{\alpha\beta}\partial_\mu a^\mu\ .
	\ee
Hence $J^{\alpha\beta}(x)$ satisfies the generalized conservation equation
\be\label{cont-fratt'}
	\partial_\alpha\partial_\beta J^{\alpha\beta}=0\ ,
	\ee
which holds off-shell and coincides with the fractonic one \cite{Bertolini:2022ijb}. We shall see shortly that the theory described by $S_{inv}$ \eqref{Sinv} for a generic tensor field $a_{\mu\nu}(x)$ displays indeed a fractonic sector. This is not surprising, since the transformation \eqref{dxi} contains the longitudinal diffeomorphisms \eqref{dfract}, which characterize the covariant description of fractons done in \cite{Bertolini:2022ijb,Blasi:2022mbl,Bertolini:2023juh}. It is easy to verify that, using the EoM \eqref{eom a} and \eqref{eom h}, the current $j^{\mu\nu}(x)$ \eqref{j} is conserved
	\be\label{cont-j}
	\partial_\mu j^{\mu\nu}=\partial_\mu J^{\mu\nu}-\frac{1}{2}\epsilon^{\mu\nu\alpha}\partial_\mu J_\alpha=0\ .
	\ee
Notice that \eqref{cont-j} represents a standard on-shell conservation equation, and that the currents $J^{\alpha\beta}(x)$ and $J^{\alpha}(x)$ are not separately conserved
	\begin{align}
	\partial_\beta J^{\alpha\beta}&=
\epsilon^{\alpha\mu\nu}\partial_\mu\left(\partial^\beta h_{\nu\beta}+\epsilon_{\nu\rho\lambda}\partial^\rho a^\lambda\right)\neq0\\
	\partial_\alpha J^\alpha&=\frac1 2 \partial_\alpha\left(\partial^\alpha h-\partial_\alpha\partial_\beta h^{\alpha\beta}\right)\neq0\ ,\label{std cont}
	\end{align}
but they do only if combined as in \eqref{j}, which therefore is a true matter current. Indeed we can extend the electromagnetic-like behaviors \eqref{Jab=E} and \eqref{Ja0=B} also to the generalized non-symmetric current $j^{\mu\nu}(x)$ \eqref{j}. From  \eqref{Jab=E}-\eqref{Ja=E} we find
\begin{align}
j^{a0}&=\mathcal B^a_-
\label{ja0}\\
j^{0a}&=\mathcal B^a_+\label{j0a}\\
j^{ab}&=J^{ab}-\frac1 2 \epsilon^{0ab}J_0=\frac1 2 \left(\epsilon^{0am}E^b_{\ m}+\epsilon^{0bm}E^a_{\ m}+\epsilon^{0ab}E^m_{\ m}\right)\label{jab=Etot}\ ,
\end{align}
where we identified  electric and magnetic fields which generalize the ones given in \eqref{tensorE} and \eqref{vectorB}
	\begin{align}
	\mathcal E_{ab}&\equiv G_{ab0}=\tilde G_{ab0}+\frac{1}{2}\eta_{ab}G^m_{\ m0}=\tilde E_{ab}+\frac{1}{2}\eta_{ab}\mathcal E^m_{\ m}\label{Etot}\\
	\mathcal B^a_\pm&\equiv B^a\pm\frac{1}{2}\epsilon^{0ab}G^\mu_{\ \mu b}\ .\label{Bpm}
	\end{align}
We finally observe that \eqref{Jab=E} and \eqref{jab=Etot} look like higher rank Hall-like Ohm's Laws, since we can write
	\begin{align}
	J^{ab}&=\tilde\sigma^{abmn}\tilde E_{mn}\\
	j^{ab}&=\sigma^{abmn}\mathcal E_{mn}\ ,\label{ohm1}
	\end{align}
with 
	\begin{align}
	\tilde\sigma^{abmn}&\equiv\frac{1}{4}\left(\epsilon^{0am}\eta^{bn}+\epsilon^{0bm}\eta^{an}+\epsilon^{0an}\eta^{bm}+\epsilon^{0bn}\eta^{am}\right)\label{tsigma}\\
	\sigma^{abmn}&\equiv\tilde\sigma^{abmn}+\frac{1}{2}\epsilon^{0ab}\eta^{mn}\ ,\label{sigma}
	\end{align}
which may be identified with generalized tensor conductivities.

\subsection{Emergence of fractons}

From \eqref{cont-fratt'}, and \eqref{J00=0}, and defining
	\be
	\rho\equiv 2\partial_aJ^{0a}\ ,
	\ee
we get the continuity equation which characterizes fracton theories \cite{Bertolini:2022ijb,Bertolini:2023sqa,Pretko:2016lgv,Bertolini:2024yur,Pena-Benitez:2023aat,Seiberg:2020bhn,Gromov:2017vir,Bidussi:2021nmp,Gromov:2022cxa}
	\be\label{fc}
	\partial_0\rho+\partial_a\partial_bJ^{ab}=0\ .
	\ee
Fractonic nature is encoded in this continuity equation through conservation relations that constrain the motion of the quasiparticles. In particular \eqref{fc} implies the conservation of charge, dipole and of the trace of the quadrupole momentum as follows
	\begin{align}
	\partial_0\int dV\rho=&-\int dV\partial_a\partial_bJ^{ab}=0\label{charge cons}\\
	\partial_0\int dVx^i\rho=&-\int dVx^i\partial_a\partial_bJ^{ab}=\int dV\partial_aJ^{ai}=0\label{dipole cons}\\
	\partial_0\int dVx^2\rho=&-\int dVx^2\partial_a\partial_bJ^{ab}=-2\int dVJ^m_{\ m}=0\ ,\label{quadrupole}
	\end{align}
where we  used the fact that, due to \eqref{diva=0} and \eqref{J00=0}, we have that the current \eqref{Jab=E} is spatially traceless
	\be\label{TrJ=0}
	J^m_{\ m}=0\ ,
	\ee
and we also observed that the total dipole momentum is
	\be\label{D}
	D^a\equiv\int dV\,d^a=\int dV\,x^a\rho=-2\int dV\,J^{0a}\ .
	\ee
A crucial remark is that the conservation equations \eqref{charge cons}, \eqref{quadrupole}, and in particular the dipole conservation \eqref{dipole cons}, imply that single charges  cannot move in isolation and are thus identifiable as fractons. On the contrary, the dipoles $d^i(x)$ are only partially constrained by the quadrupole conservation \eqref{quadrupole} to move along their transverse direction, $i.e.$ on one dimension, which allows to identify them as \textit{lineons} \cite{Shirley:2018nhn}. These mobility properties characterize the traceless scalar charge theory of fractons described in \cite{Pretko:2016lgv,Pretko:2016kxt}. We can therefore make the following identifications
	\begin{align}
	\rho&\equiv 2\partial_aJ^{0a}&&\mbox{\hspace{-130px}}\to\mbox{ fractonic charge}\label{fracton charge}\\
	 d^a&\equiv-2 J^{0a}&&\mbox{\hspace{-130px}}\to\mbox{ dipole momentum density}\label{fracton current}\\
	  &\qquad J^{ab}&&\mbox{\hspace{-130px}}\to\mbox{ dipole current}\ .\label{dipole current}
	\end{align}
Thus from the higher rank conserved current \eqref{cont-fratt'}, fractonic features emerge through the continuity equation \eqref{fc}. We now analyze the  standard conserved current \eqref{cont-j}, which we analyze separating space and time indices
	\bi
	\item $\pmb{\nu=0}$
	\be\label{djm0}
	\partial_mj^{m0}=\partial_mJ^{0m}+\frac{1}{2}\epsilon^{0ma}\partial_mJ_a=\frac{1}{2}\left(\rho+\epsilon^{0ab}\partial_aJ_b\right)=0
\ ,
	\ee
where we used \eqref{fracton charge}. This equation tells us that 
	\be\label{J=-ro}
	\epsilon^{0ab}\partial_aJ_b=-\rho\ ,
	\ee
from which we observe that the current $J_a(x)$ can be decomposed into two contributions
	\be\label{J=0+1}
	J_a=J^{(0)}_a+J^{(1)}_a\ ,
	\ee
such that
	\begin{align}
	&&\epsilon^{0ab}\partial_aJ^{(0)}_b&=0\mbox{\hspace{-30px}}&\Rightarrow&&J^{(0)}_a&=\partial_a f&&\label{deff}\\
	&&\epsilon^{0ab}\partial_a J^{(1)}_b&=-\rho\mbox{\hspace{-30px}}&\Rightarrow&&J^{(1)}_a&=\epsilon_{0ab}d^b\label{mJ=rho}\ ,&&
	\end{align}
where $f(x)$ is a generic scalar function, and in \eqref{mJ=rho} we used \eqref{D}. Therefore \eqref{mJ=rho} implies that a part of the vector current $J_a(x)$, represented by $J^{(1)}_a(x)$, is related to the fracton charge \eqref{fracton charge} and can be associated to dipolar behaviours through \eqref{D}, while $J^{(0)}_a(x)$ depends on a generic function  $f(x)$, and thus is not necessarily related to dipoles.
\item $\pmb{\nu=n}$
	\be\label{djmn}
	-\tfrac{1}{2}\partial_0\left(d^n+\epsilon^{0na}J_a\right)+\partial_m\left(J^{mn}-\tfrac{1}{2}\epsilon^{mn0}J_0\right)=0\ ,
	\ee	
where we used the dipole identification \eqref{fracton current}. It is a standard (meaning with one derivative) continuity equation
	\be\label{vector cont}
	\partial_0\omega^n+\partial_mj^{mn}=0\ ,
	\ee
 which involves a vectorial quantity 
	\be\label{omega}
	\omega ^n\equiv-\frac{d^n+\epsilon^{0na}J_a}{2}=-d^n-\frac1 2 \epsilon^{0na}\partial_af=j^{0n}\ .
	\ee
In the second equality we used \eqref{J=0+1} in order to highlight that the vector charge $\omega^i(x)$ is composed of both the dipole contribution $d^i(x)$ and the non-dipole one $f(x)$. To better understand the physical nature of the vector charge $\omega^a(x)$ \eqref{omega}, we can analyze its conserved quantities as done in \eqref{charge cons}-\eqref{quadrupole}. 
\begin{enumerate}
	\item  From \eqref{vector cont} we see that the total vector charge is conserved
	\be\label{cons-ch}
	\partial_0 \int dV\,\omega ^i=0\ .
	\ee
	\item Another conserved quantity is given by
		\be\label{cons-xch}
		\partial_0\int dV\,x_n\omega^n=\int dV\left(J^m_{\;m}-\frac1 2\eta_{mn}\epsilon^{0mn}J_0\right)=0\ .
		\ee
where we used \eqref{TrJ=0}. This conservation gives a constraint on the motion of the vector charge $\omega^i(x)$, as the quadrupole conservation \eqref{quadrupole} constrains the movement of the dipole $d^i(x)$ \eqref{fracton current}
	\be
	\partial_0\int dV\,x^2\rho=\partial_0\int dV\,x_id^i=0\ ,
	\ee
due to \eqref{D}. Thus the vectorial charge $\omega^a(x)$ \eqref{omega} moves only along its transverse direction, as the dipole $d^i(x)$ \eqref{fracton current}.
\end{enumerate}
Notice that if $J_0(x)=0$, from the continuity equation \eqref{vector cont} two additional conserved quantities can be recovered.
	\begin{enumerate}
	\item[3.] An angular momentum-like conservation	
		\be
		\partial_0\int dV\,\epsilon_{0bn}x^b\omega ^n=0\label{conserv1}\ ,
		\ee
		which through \eqref{omega} and \eqref{D} is directly related to the scalar function $f(x)$ in \eqref{deff}
		\be\label{f=angmom}
		\int dV\,\epsilon_{0bn}x^b\omega ^n=-\frac{1}{2}\int dVx^a\partial_af=\int dV\, f\ .
		\ee
	\item[4.] The conservation of a vector
		\be
		\partial_0\int dV\left(x^ax_n\omega ^n-\tfrac1 2x^2\omega ^a\right)=0\ ,\label{conserv2}
		\ee
where again we used \eqref{TrJ=0}.
	\end{enumerate}
Remarkably, the conservation relations  \eqref{cons-ch}, \eqref{cons-xch}, \eqref{conserv1} and \eqref{conserv2} exactly coincide with those identifying the traceless vector charge theory of fractons \cite{Pretko:2016lgv,Pretko:2016kxt}. 
In that case the vectorial charge can be identified as $\omega ^i(x)$ \eqref{omega}, whose mobility is completely constrained, making it a true fracton. The traceless vector charge theory  has a continuity equation \cite{Pretko:2016kxt} that indeed coincides with our \eqref{djmn} when setting $J_0(x)=0$, $i.e.$ the space-derivative  depends on the traceless symmetric contribution $J^{mn}(x)$ \eqref{dipole current}. The tracelessness property of the traceless vector charge fracton model is related to the vanishing of the trace of a rank-2 symmetric electric tensor \cite{Pretko:2016lgv,Pretko:2016kxt}. 
The same happens here when noticing that from the EoM of $a_0(x)$ \eqref{J0=B}, the $J_0(x)$ current is associated to the trace of the electric field $\mathcal E_{ab}(x)$ \eqref{Etot}
	\be
	J^0=\frac{\delta S_{inv}}{\delta a_0}=-G_m^{\ m0}=\mathcal E^m_{\ m}\ ,
	\ee
and thus when $J_0(x)=0$ the generalized electric field \eqref{Etot} is traceless. 
\ei
Therefore the analysis of the two continuity equations \eqref{cont-fratt'} and \eqref{vector cont}, leads us to claim that the theory represented by $S_{inv}$ \eqref{Sinv} coupled to the matter term $S_J$ \eqref{SJ} has fractonic features, since it describes charges with restricted motion. In particular, a novelty is that we have a vectorial quantity $\omega ^i(x)$ \eqref{omega}, whose existence is a consequence of the introduction of the antisymmetric contribution $a_\mu(x)$. The vector $\omega^i(x)$ is characterized by a partially constrained motion, which could be related to a new fractonic behavior. From \eqref{djm0}, and from the identification of the fractonic charge $\rho(x)$ \eqref{fracton charge} it follows that the magnetic vector fields \eqref{ja0} and \eqref{j0a} satisfy
	\be\label{dB-=0}
	\partial_a\mathcal B^a_-=0\quad
;\quad
\partial_aB^a=\tfrac1 2\rho\quad
;\quad\partial_a\mathcal B_+^a=\rho\ .
	\ee
Thus the magnetic vector fields $B^a(x)$ and $\mathcal B_+^a(x)$ have non-vanishing divergences, related to the fractonic charge $\rho(x)$ \eqref{fracton charge}.
Thanks to the identification of the conserved charges \eqref{fracton current} and \eqref{omega} and their fractonic features, we can finally interpret the electromagnetic-like equations \eqref{Ja0=B}, \eqref{j0a}, \eqref{Jab=E} and \eqref{jab=Etot}  of the previous Subsection in terms of a Hall-like generalization. In particular \eqref{Ja0=B} and \eqref{j0a} are the higher rank extention of the standard flux-attachment relation of Chern-Simons theory \cite{Dunne:1998qy,tong} that arises from the EoM of the theory coupled to matter
	\be
	J^0=\frac{\delta S_{CS}}{\delta A_0}\quad\Rightarrow\quad\rho_{em}=B\ .\label{CSflux}
	\ee
The effect of the relation \eqref{CSflux} is to attach the flux of the scalar magnetic field $B(x)$ to the matter charge density $\rho_{em}(x)$ in such a way that the flux follows the charge density \cite{Dunne:1998qy,tong}. In our case Eqs. \eqref{Ja0=B} and \eqref{j0a}, involve the vectorial charges $d^i(x)$ \eqref{fracton current}, $\omega^i(x)$ \eqref{omega} and the vectorial magnetic fields $B^a(x)$ and $\mathcal B_+^a(x)$, and are thus vectorial extensions of \eqref{CSflux}
	\begin{align}
	d^a&=-2B^a\label{d=B}\\
	\omega^a&=\mathcal B^a_+\ ,\label{w=B}
	\end{align}
with \eqref{ja0} being simply a combination of the two
	\be
	d^a+\omega^a=-\mathcal B_-^a\ .
	\ee
The higher rank relations \eqref{d=B}, \eqref{w=B} thus imply that the magnetic fluxes of $B^a(x)$ and $\mathcal B_+^a(x)$ follow the (vectorial) charges wherever they go. The difference from the Chern-Simons case \eqref{CSflux} is that this theory has fractonic features. Indeed, since we saw from the constraints \eqref{charge cons}-\eqref{quadrupole} and \eqref{cons-ch}, \eqref{cons-xch} that both $d^i(x)$ and $\omega^i(x)$ move in one spatial dimension when $J_0(x)\neq0$, so do the fluxes, while if $J_0(x)=0$, the additional constraints \eqref{conserv1}, \eqref{conserv2} arise for $\omega^i(x)$, implying that the flux of $\mathcal B_+^a(x)$ is localized and immobile, as its associated charge. Furthermore, from \eqref{Jab=E} and \eqref{jab=Etot} we observed the Hall-like conductivities \eqref{tsigma} and \eqref{sigma},
 which we can now relate to the transport of dipoles $d^i(x)$ \eqref{fracton current} \cite{Prem:2017kxc,Bertolini:2024yur,Pretko:2017xar}, and to the transport of the vector charge $\omega^a(x)$, respectively. 
Notice also that from \eqref{jab=Etot}, if $J_0(x)=0$, we have
	\be
	j^{ab}=J^{ab}=\frac1 2 \left(\epsilon^{0am}\tilde E^b_{\ m}+\epsilon^{0bm}\tilde E^a_{\ m}\right)=\tilde\sigma^{abmn}\tilde E_{mn}\ ,
	\ee
which is a confirmation of the fact that if $J_0(x)=0$, the theory only contains a traceless electric tensor field $\tilde E^{ab}(x)$ \eqref{tensorE}. There is thus only one Hall-like conductivity $\tilde\sigma^{abmn}$, which is not surprising since the case $J_0(x)=0$ represents the traceless vector charge theory for fractons \cite{Pretko:2016lgv,Pretko:2016kxt} for which the vector charge $\omega^a(x)$ is completely immobile. The only remaining current is  $J^{ab}(x)$ which is indeed the one associated to the mobile dipoles $d^a(x)$ \eqref{fracton current}.

\section{Summary and conclusions} 

The absence of any ability to move lies at the heart of the definition of a fracton quasiparticle \cite{Nandkishore:2018sel,Pretko:2020cko}. These phases of matter share connections with many other areas of physics, such as elasticity \cite{Pretko:2017kvd,Pretko:2019omh}, hydrodynamics \cite{Glodkowski:2024ova,Gromov:2020yoc}, gravity \cite{Pretko:2017fbf}, and Carrollian theories \cite{Figueroa-OFarrill:2023vbj,Figueroa-OFarrill:2023qty}. The common features being immobility, manifesting in the form of dipole conservation, and the presence of a rank-2 symmetric tensor field. Recently it has been shown \cite{Bertolini:2024yur} that a covariant 3D Chern-Simons-like theory described by a rank-2 symmetric tensor field $h_{\mu\nu}(x)$ displays fractonic properties through a generalized planar electromagnetism and dipole conservation. Three dimensional fracton theories are of particular interest in physical systems describing, for instance, elastic media \cite{Pretko:2017kvd,Pretko:2019omh}. In this context, in order to take into account non-standard behaviours, generalizations involving non-symmetric fields have been considered \cite{Pretko:2019omh,Gromov:2019waa,Surowka:2021ved}. Taking insights from these examples, in this paper we studied the most general covariant action described by a rank-2 non-symmetric field, which transforms under \eqref{dxi} and includes the covariant fractonic symmetry through \eqref{dfract}. From the gauge field theory point of view, our analysis shows that the 3D theory described by the action \eqref{Sinv}
	\bi
	\item is ``quasi-topological'', in the sense that, despite the fact that the action $S_{inv}$ \eqref{Sinv} depends on the spacetime metric, the energy-momentum tensor vanishes on-shell, which closely reminds what happens in topological quantum field theories, whose invariant actions are metric independent and, consequently, have vanishing off-shell energy momentum tensor.
	\item displays a peculiar gauge fixing structure. Indeed the propagators display poles in the gauge parameters \eqref{poli}, which opens the possibility of non-intuitive gauge fixing terms. For instance gauge fixing conditions concerning only the traceless-symmetric part of the gauge field $a_{\mu\nu}(x)$ \eqref{Sgf01/3}, or combinations of the trace and of the antisymmetric components \eqref{gf0kl} are good ones, contrarily to what one might expect. 
	\item has three DoF. Therefore, compared to the symmetric theory of \cite{Bertolini:2024yur} which has two DoF, we deduce that the additional DoF is carried by the antisymmetric part of the rank-2 tensor field $a_{\mu\nu}(x)$. As we shall discuss, this has a physical interpretation.
	\ei
The  theory described by the action \eqref{Sinv} features a form of generalized electromagnetism, with additional fields with respect to its symmetric Chern-Simons-like counterpart \cite{Bertolini:2024yur}. For instance the 3D theory of a symmetric tensor \cite{Bertolini:2024yur} is traceless, while in the case of a generic tensor, we find that, due to a coupling with the antisymmetric contribution, the trace $h(x)$ plays the role of a multiplier for a kind of ``internal'' gauge fixing condition \eqref{diva=0}. 
This reflects in the electromagnetic interpretation: from the EoM of the symmetric field $h_{\mu\nu}$ \eqref{eomh}, a traceless invariant electric field $\tilde E^{ij}(x)$ \eqref{tracelessE} can be identified, together with a vectorial magnetic field $B^a(x)$ \eqref{vectorB}, in agreement with \cite{Bertolini:2024yur}. However the presence of an antisymmetric component leads to the additional EoM \eqref{eom a} and allows to define also a traceful electric tensor field $\mathcal E^{ij}(x)$ \eqref{Etot} and magnetic vector fields $\mathcal B^a_\pm(x)$ \eqref{Bpm}, which are new invariant quantities. These identifications are a first sign that the theory is related to fractons,  which are characterized by an electromagnetic-like behaviour \cite{Bertolini:2022ijb,Pretko:2016lgv}. In particular, the theory described by the action \eqref{Sinv} can be viewed as a new higher rank electromagnetic analogous to the standard Chern-Simons theory \cite{Dunne:1998qy,tong}. As in the standard Chern-Simons theory, a physical interpretation is possible only  when the matter term $S_J$ \eqref{SJ} is introduced. 
The main consequence is the existence of the continuity equation \eqref{fc}, which encodes the dipole conservation \eqref{dipole cons}, which is the basic property of the fracton behaviour \cite{Nandkishore:2018sel,Pretko:2020cko}. This therefore confirms that the theory represented by the action \eqref{Sinv} describes indeed a fracton model. Both mobile and immobile quasiparticles can be identified from the continuity equations \eqref{fc} and \eqref{vector cont}, which are consequence of the conserved current equations \eqref{cont-fratt'} and \eqref{cont-j} related to the defining symmetry \eqref{dxi} and the underlying fracton symmetry \eqref{dfract}. In particular we can  identify the fractonic immobile charge $\rho(x)$ \eqref{fracton charge}, the dipole density $d^i(x)$ \eqref{fracton current}, and the vector charge $\omega^i(x)$ \eqref{omega}. 
Notice that the last one, $\omega^i(x)$, differs from the dipole $d^i(x)$ by the curl of a scalar quantity $\epsilon^{0ij}\partial_jf(x)$ \eqref{omega}. From \eqref{f=angmom} it can be seen that this scalar function $f(x)$ is a planar angular momentum, and it is related to the matter source by $J^a(x)$ \eqref{deff}. It is thus a direct consequence of the presence of the antisymmetric vectorial component $a_{\mu}(x)$. From the analysis of the continuity equations \eqref{fc} and \eqref{vector cont}, we can update Table \ref{Table2} and summarize that the theory defined by the symmetry \eqref{dxi} and the non-symmetric rank-2 tensor $a_{\mu\nu}(x)$ \eqref{a} contains both 
	\bi
	\item a traceless scalar charge theory of fractons \cite{Pretko:2016lgv,Pretko:2016kxt} for an immobile charge density $\rho(x)$ \eqref{fracton charge} (\textbf{fracton}) and a dipole density $d^i(x)$ \eqref{fracton current} moving in one dimension (\textbf{lineon})
	\item a vector-like theory for $\omega^i(x)$ \eqref{omega}, moving in one dimension (\textbf{lineon}), which reduces to the traceless vector charge theory of fractons \cite{Pretko:2016lgv,Pretko:2016kxt} when $J_0(x)=0$ for which $\omega^a(x)$ becomes immobile (\textbf{fracton}).
	\ei
It is a remarkable fact that we find the typical fractonic properties, which can be summarized in the existence of a generalized electromagnetism and the presence of fracton quasiparticles with limited mobility, without restricting to the longitudinal diffeomorphisms \eqref{dfract}, but keeping the full symmetry \eqref{dxi} which contains the infinitesimal diffeomorphisms \eqref{diff}. Interestingly, the identification of the charges \eqref{fracton charge}, \eqref{fracton current}, \eqref{omega} and the higher rank electromagnetic fields \eqref{tracelessE}, \eqref{Etot}, \eqref{vectorB}, \eqref{Bpm} allows for a Hall-like interpretation of this model. The dipolar Hall behaviour already observed in the symmetric theory \cite{Bertolini:2024yur} can be extended to the non-symmetric theory \eqref{Sinv} and to its additional vectorial charge $\omega^i(x)$ \eqref{omega}. In particular its transport Ohm's-like law \eqref{ohm1} is related to the electric tensor field $\mathcal E^{ij}(x)$ \eqref{Etot} with conductivity $\sigma^{ijmn}$ \eqref{sigma}. Analogously, the vectorial flux-attachment relation \eqref{w=B} extends the possibility of building a fracton(lineon)/vortex duality for dipoles \cite{Huang:2023zhp} also to the $\omega^i(x)$ charge, in relation with the flux of the magnetic vector field $\mathcal B_+^a(x)$.  It is also interesting to look at this 3D model from the point of view of the fracton-elasticity duality \cite{Pretko:2017kvd,Pretko:2019omh}. In particular, there exists a similarity with the so-called ``torsional Chern-Simons theory'' studied in \cite{Gromov:2017vir} in relation with fractons, where the theory can be reconduced to Pretko's symmetric non-covariant model \cite{Pretko:2017xar} under the assumption of area preserving diffeomorphisms, while the full theory could be related to torsional viscosity effects. In this respect, it is interesting to notice that the same action of \cite{Gromov:2017vir} was previously studied in \cite{Hughes:2011hv}, when fracton were not developed yet. There, the action depends both on symmetric and antisymmetric fields, and coincides with ours when taking into account the fact that in 3D an antisymmetric field can be written as a vector through \eqref{a}.  In \cite{Hughes:2011hv} these contributions are interpreted in the context of ``Cosserat elasticity'' \cite{ERINGEN1967191,Hehl:2007bn} as related to the stress tensor (symmetric) and to the rotational DoF (antisymmetric). It thus seems that the 3D covariant theory of fractons for a rank-2 non-symmetric field could be interpreted, when analyzed in the elasticity-duality context, as the dual of Cosserat elasticity, where the antisymmetric component plays the role of the rotational DoF, which is therefore worth to further investigate. Additionally, both theories in \cite{Gromov:2017vir} and \cite{Hughes:2011hv} are related to torsional effects, which implies a possible underlying connection between the non-symmetric tensor $a_{\mu\nu}(x)$, and thus of its fractonic theory, with torsion. Finally the ``quasi-topological'' nature of the model could hint towards the possibility of having an interesting boundary physics, as it happens for topological theories \cite{Amoretti:2014iza,Bertolini:2022sao,Bertolini:2023sqa,Bertolini:2021iku,Bertolini:2023wie}.

\section*{Acknowledgments}
We thank Alessio Caddeo, Matteo Carrega, Silvia Fasce, Daniele Musso, Giandomenico Palumbo and Daniel Sacco Shaikh for enlightening discussions. This work has been partially supported by the INFN Scientific Initiative GSS: ``Gauge Theory, Strings and Supergravity''. 
\appendix

\section{Propagators}\label{App-Prop}

The gauge fixed  action $S_{tot}$ \eqref{Stot} is 
	\be\label{app-Stot}
	S_{tot}=\int d^3x\left[\epsilon^{\mu\nu\rho}h_{\mu}^{\;\lambda}\partial_\nu h_{\rho\lambda}-\epsilon^{\mu\nu\rho}a_\mu\partial_\nu a_\rho+2\left(h_{\mu\nu}\partial^\mu a^\nu-h\partial_\mu a^\mu\right)+b^\mu\left(\kappa_1\partial^\nu h_{\mu\nu}+\kappa_2\partial_\mu h+\kappa_3\epsilon_{\mu\nu\rho}\partial^\nu a^\rho\right)\right]\ ,
	\ee
which, in momentum space, reads
		\begin{align}
		S_{tot}=\int d^3p&\left[-\frac{1}{4}\hat h_{\mu\nu}\mathcal{E}^{\mu\nu,\alpha\beta}\hat h_{\alpha\beta}-\hat h_{\mu\nu}n^{\mu\nu,\alpha}\hat a_{\alpha}+i\hat a_{\mu}\epsilon^{\mu\nu\alpha}p_\nu\hat a_\alpha+\frac{1}{2}\hat h_{\mu\nu}n^{\mu\nu,\alpha}_{(\kappa_1,\kappa_2)}\hat b_{\alpha} -i\kappa_3 \hat a_\mu\epsilon^{\mu\nu\alpha}p_\nu\hat b_\alpha\right]\nonumber\\
		=\int d^3p&\hat\Phi_A M^{AB}\hat\Phi_B\ ,
		\end{align}
where
	\begin{empheq}{align}
	\mathcal{E}^{\mu\nu,\alpha\beta}&\equiv ip_\lambda\left(\epsilon^{\mu\lambda\alpha}\eta^{\nu\beta}+\epsilon^{\mu\lambda\beta}\eta^{\nu\alpha}+\epsilon^{\nu\lambda\alpha}\eta^{\mu\beta}+\epsilon^{\nu\lambda\beta}\eta^{\mu\alpha}\right)\\
	n^{\mu\nu,\alpha}&\equiv i\left(p^\mu\eta^{\nu\alpha}+p^\nu\eta^{\mu\alpha}-2p^{\alpha}\eta^{\mu\nu}\right)\\
	n^{\mu\nu,\alpha}_{(\kappa_1,\kappa_2)}&\equiv i\left[\kappa_1\left(p^\mu\eta^{\nu\alpha}+p^\nu\eta^{\mu\alpha}\right)+2\kappa_2p^{\alpha}\eta^{\mu\nu}\right]\ ,
	\end{empheq}
and
	\be
	\hat\Phi_A\equiv(h_{\mu\nu},a_{\mu},b_{\mu})\quad;\quad A=\{\mu\nu,\mu,\mu\},\ B=\{\alpha\beta,\alpha,\beta\}\ ,
	\ee
	\be
	M^{AB}=\left(
		\begin{array}{ccc}
		-\frac{1}{4}\mathcal{E}^{\mu\nu,\alpha\beta}&-\frac{1}{2}n^{\mu\nu,\alpha}&\frac{1}{4}n_{(\kappa_1,\kappa_2)}^{\mu\nu,\alpha}\\
		-\frac{1}{2}n^{*\alpha\beta,\mu}&i\epsilon^{\mu\lambda\alpha}p_\lambda& -\frac{i}{2}\kappa_3\epsilon^{\mu\lambda\alpha}p_\lambda\\
		\frac{1}{4}n_{(\kappa_1,\kappa_2)}^{*\alpha\beta,\mu}& -\frac{i}{2}\kappa_3\epsilon^{\mu\lambda\alpha}p_\lambda&0\\
		\end{array}\right)\ .
	\ee
Notice that $n^{\mu\nu,\alpha}_{(\kappa_1,-\kappa_1)}=\kappa_1 n^{\mu\nu,\alpha}$. The matrix propagator $D_{AB}$ is the solution of  
	\be\label{app-prop eq}
	M^{AC}D_{CA}=I^A_{\;B}\ ,
	\ee
where
	\be
	D^{BC}\equiv\left(
		\begin{array}{ccc}
		{H}_{\alpha\beta,\rho\sigma}&{H}_{\alpha\beta,\rho}&{\mathcal{H}}_{\alpha\beta,\rho}\\
		{H}^{*}_{\rho\sigma,\alpha}&A_{\alpha\rho}&\mathcal{A}_{\alpha\rho}\\
		{\mathcal{H}}^*_{\rho\sigma,\alpha}&\mathcal{A}_{\alpha\rho}&B_{\alpha\rho}\\
		\end{array}\right)\ ,
	\ee
and the following expansion hold \cite{Bertolini:2024yur,Blasi:2017pkk,Maggiore:2019wie,Bertolini:2020hgr}
	\begin{empheq}{align}
	{H}_{\alpha\beta,\rho\sigma}&\equiv c_0\mathcal{I}_{\alpha\beta,\rho\sigma}+c_1\left(\eta_{\alpha\rho}p_\beta p_\sigma+\eta_{\alpha\sigma}p_\beta p_\rho+\eta_{\beta\rho}p_\alpha p_\sigma+\eta_{\beta\sigma}p_\alpha p_\rho\right)+\nonumber\\
	&\quad+c_2\left(\eta_{\alpha\beta}p_\rho p_\sigma+\eta_{\rho\sigma} p_\alpha p_\beta\right)+c_3\eta_{\alpha\beta}\eta_{\rho\sigma}+c_4p_\alpha p_\beta p_\rho p_\sigma+c_5\mathcal{E}_{\alpha\beta,\rho\sigma}+\label{app-Delta}\\
	&\quad+ic_6p^\lambda\left(\epsilon_{\alpha\lambda\rho}p_\sigma p_\beta+\epsilon_{\alpha\lambda\sigma}p_\beta p_\rho+\epsilon_{\beta\lambda\rho}p_\alpha p_\sigma+\epsilon_{\beta\lambda\sigma}p_\alpha p_\rho\right)\nonumber\\
	{H}_{\alpha\beta,\rho}&\equiv ic_7\eta_{\alpha\beta}p_{\rho}+ic_8p_\alpha p_\beta p_{\rho}+c_9p^\lambda
\left(\epsilon_{\alpha\lambda\rho}p_\beta+\epsilon_{\beta\lambda\rho}p_\alpha\right)+ic_{10}\left(\eta_{\alpha\rho}p_\beta+\eta_{\beta\rho}p_\alpha\right)\label{app-D}\\
	{\mathcal{H}}_{\alpha\beta,\rho}&\equiv ic_{11}\eta_{\alpha\beta}p_{\rho}+ic_{12}p_\alpha p_\beta p_{\rho}+c_{13}p^\lambda
\left(\epsilon_{\alpha\lambda\rho}p_\beta+\epsilon_{\beta\lambda\rho}p_\alpha\right)+ic_{14}\left(\eta_{\alpha\rho}p_\beta+\eta_{\beta\rho}p_\alpha\right)\label{app-d}\\
	A_{\alpha\rho}&\equiv c_{15}\eta_{\alpha\rho}+c_{16}p_{\alpha} p_{\rho}+ic_{17}p^\lambda\epsilon_{\alpha\lambda\rho}\label{app-A}\\
	\mathcal{A}_{\alpha\rho}&\equiv c_{18}\eta_{\alpha\rho}+c_{19}p_{\alpha} p_{\rho}+ic_{20}p^\lambda\epsilon_{\alpha\lambda\rho}\label{app-McA}\\
	B_{\alpha\rho}&\equiv c_{21}\eta_{\alpha\rho}+c_{22}p_{\alpha} p_{\rho}+ic_{23}p^\lambda\epsilon_{\alpha\lambda\rho}\ ,\label{app-B}
	\end{empheq}
which depends on $c_i$, $i=0,\ldots,23$ constant parameters to be determined.
\allowdisplaybreaks
From \eqref{app-prop eq} we get nine equations
	\begin{empheq}{align}
	-\frac{1}{4}\mathcal{E}^{\mu\nu,\alpha\beta}{H}_{\alpha\beta,\rho\sigma}-\frac{1}{2}n^{\mu\nu,\alpha}{H}^*_{\rho\sigma,\alpha}+\frac{1}{4}n^{\mu\nu,\alpha}_{(\kappa_1,\kappa_2)}{\mathcal{H}}^*_{\rho\sigma,\alpha}&=\mathcal{I}^{\mu\nu}_{\;\rho\sigma}\label{app-prop1}\\
	-\frac{1}{4}\mathcal{E}^{\mu\nu,\alpha\beta}{H}_{\alpha\beta,\rho}-\frac{1}{2}n^{\mu\nu,\alpha}A_{\alpha\rho}+\frac{1}{4}n^{\mu\nu,\alpha}_{(\kappa_1,\kappa_2)}\mathcal{A}_{\alpha\rho}&=0\label{app-prop2}\\
	-\frac{1}{4}\mathcal{E}^{\mu\nu,\alpha\beta}{\mathcal{H}}_{\alpha\beta,\rho}-\frac{1}{2}n^{\mu\nu,\alpha}\mathcal{A}_{\alpha\rho}+\frac{1}{4}n^{\mu\nu,\alpha}_{(\kappa_1,\kappa_2)}B_{\alpha\rho}&=0\label{app-prop3}\\[5px]
	-\frac{1}{2}n^{*\alpha\beta,\mu}{H}_{\alpha\beta,\rho\sigma}+ip_\lambda\epsilon^{\mu\lambda\alpha}{H}^*_{\rho\sigma,\alpha} -\frac{i}{2}\kappa_3p_\lambda\epsilon^{\mu\lambda\alpha}{\mathcal{H}}^*_{\rho\sigma,\alpha}&=0\label{app-prop4}\\
	-\frac{1}{2}n^{*\alpha\beta,\mu}{H}_{\alpha\beta,\rho}+ip_\lambda\epsilon^{\mu\lambda\alpha}A_{\alpha\rho} -\frac{i}{2}\kappa_3p_\lambda\epsilon^{\mu\lambda\alpha}\mathcal{A}_{\alpha\rho}&=\delta^{\mu}_{\rho}\label{app-prop5}\\
	-\frac{1}{2}n^{*\alpha\beta,\mu}{\mathcal{H}}_{\alpha\beta,\rho}+ip_\lambda\epsilon^{\mu\lambda\alpha}\mathcal{A}_{\alpha\rho} -\frac{i}{2}\kappa_3p_\lambda\epsilon^{\mu\lambda\alpha}B_{\alpha\rho}&=0\label{app-prop6}\\[5px]
		\frac{1}{4}n_{(\kappa_1,\kappa_2)}^{*\alpha\beta,\mu}{H}_{\alpha\beta,\rho\sigma} -\frac{i}{2}\kappa_3p_\lambda\epsilon^{\mu\lambda\alpha}{H}^*_{\rho\sigma,\alpha}&=0\label{app-prop7}\\
		\frac{1}{4}n_{(\kappa_1,\kappa_2)}^{*\alpha\beta,\mu}{H}_{\alpha\beta,\rho} -\frac{i}{2}\kappa_3p_\lambda\epsilon^{\mu\lambda\alpha}A_{\alpha\rho}&=0\label{app-prop8}\\
		\frac{1}{4}n_{(\kappa_1,\kappa_2)}^{*\alpha\beta,\mu}{\mathcal{H}}_{\alpha\beta,\rho} -\frac{i}{2}\kappa_3p_\lambda\epsilon^{\mu\lambda\alpha}\mathcal{A}_{\alpha\rho}&=\delta^{\mu}_{\rho}\ .\label{app-prop9}
	\end{empheq}
Notice that $\mathcal{A}^*_{\rho\alpha}=c_{18}\eta_{\rho\alpha}+c_{19} p_{\rho}p_{\alpha}-ic_{20}p^\lambda\epsilon_{\rho\lambda\alpha}=c_{18}\eta_{\alpha\rho}+c_{19}p_{\alpha} p_{\rho}+ic_{20}p^\lambda\epsilon_{\alpha\lambda\rho}=\mathcal{A}_{\alpha\rho}$. 
From the above equations we get the following constraints on the $c_i$ parameters
	\begin{align}
	c_0=&0\quad\mbox{from }\eqref{app-prop1}\\
	2c_1-2c_9+\kappa_1 c_{13}=&0\\
	-3c_5+p^2c_6-c_{10}+\frac{\kappa_1}{2}c_{14}=&0\\
	c_5=&\tfrac{1}{4p^2}\\
	-2c_5+c_7+\frac{\kappa_2}{2}c_{11}=&0\\
	-2c_6-c_8+\frac{\kappa_1}{2}c_{12}=&0\\
	2c_5+p^2c_8+c_{10}+\frac{\kappa_2}{2}c_{12}+\kappa_2c_{14}=&0\\
	2c_5-c_7+\frac{\kappa_1}{2}c_{11}=&0\\[10px]
	2p^2 c_9+2c_{15}-\kappa_1 c_{18}=&0\quad\mbox{from }\eqref{app-prop2}\\
	c_9-c_{16}+\frac{\kappa_1}{2}c_{19}=&0\\
	c_{15}+p^2c_{16}+\frac{\kappa_2}{2}c_{18}+\frac{\kappa_2}{2}p^2c_{19}=&0\\
	2c_{10}+2c_{17}-\kappa_1 c_{20}=&0\\[10px]
	2p^2c_{13}+2c_{18}-\kappa_1 c_{21}=&0\quad\mbox{from }\eqref{app-prop3}\\
	c_{13}-2c_{19}+\kappa_1 c_{22}=&0\\
	c_{18}+p^2c_{19}+\frac{\kappa_2}{2}c_{21}+\frac{\kappa_2}{2}p^2c_{22}=&0\\
	2c_{14}+2c_{20}-\kappa_1 c_{23}=&0\\[10px]
	(\kappa_1+\kappa_2)(c_0+p^2c_2+3c_3)=&0\quad\mbox{from }\eqref{app-prop4}\mbox{ and }\eqref{app-prop7}\\
	(\kappa_1+\kappa_2)(4c_1+3c_2+p^2c_4)+2(\kappa_1 -\kappa_3)c_9 -\kappa_3c_{13}=&0\\
	2(\kappa_1 -\kappa_3)c_9 -\kappa_3c_{13}=&0\\
	2(\kappa_1 -\kappa_3)c_{10} -\kappa_1 \kappa_3c_{14}=&0\\[10px]
	2(\kappa_1+\kappa_2)(3c_7+p^2c_8+2c_{10})+(\kappa_1 -\kappa_3)c_{17} -\kappa_1 \kappa_3c_{20}=&0\quad\mbox{from }\eqref{app-prop5}\mbox{ and }\eqref{app-prop8}\\
	2(\kappa_1 -\kappa_3)c_{15} -\kappa_1 \kappa_3c_{18}=&0\\
	2(\kappa_1 -\kappa_3)c_{17} -\kappa_1 \kappa_3c_{20}=&-\tfrac{2\kappa_1}{p^2}\\[10px]
	2(\kappa_1 -\kappa_3)p^2c_{20} -\kappa_1 \kappa_3p^2c_{23}+4=&0\quad\mbox{from }\eqref{app-prop6}\mbox{ and }\eqref{app-prop9}\\
	2(\kappa_1+\kappa_2)(3c_{11}+p^2c_{12}+2c_{14})+2(\kappa_1 -\kappa_3)c_{20} -\kappa_1 \kappa_3c_{23}=&0\\
	2(\kappa_1 -\kappa_3)c_{18} -\kappa_1 \kappa_3c_{21}=&0\\
	\kappa_1 c_0+2\kappa_1 p^2c_1 -2\kappa_3p^2c_9=&0\quad\mbox{from }\eqref{app-prop7}\\
	2(\kappa_1+2\kappa_2)c_1+(\kappa_1+3\kappa_2)c_2+(\kappa_1+\kappa_2)p^2c_4+2\kappa_3c_9=&0\\
	\kappa_2c_0+(\kappa_1+\kappa_2)p^2c_2+(\kappa_1+3\kappa_2)c_3=&0\\
	\kappa_1 c_5+\kappa_1 p^2c_6 -\kappa_3c_{10}=&0\\[10px]
	(\kappa_1+3\kappa_2)c_7+(\kappa_1+\kappa_2)p^2c_8+(\kappa_1+2\kappa_2)c_{10} -\kappa_3c_{17}=&0\quad\mbox{from }\eqref{app-prop8}\\
	\kappa_1 p^2c_9+\kappa_3c_{15}=&0\\
	\kappa_1 c_{10}+\kappa_3c_{17}=&0\\[10px]
	\kappa_1 p^2c_{13}+\kappa_3c_{18}=&0\quad\mbox{from }\eqref{app-prop9}\\
	\kappa_1 c_{14}+\kappa_3c_{20}=&\tfrac{2}{p^2}\\
	(\kappa_1+3\kappa_2)c_{11}+(\kappa_1+\kappa_2)p^2c_{12}+(\kappa_1+2\kappa_2)c_{14} -\kappa_3c_{20}=&0\ ,
	\end{align}
which are solved by
\setlength\arraycolsep{2.5pt}
	\be\label{sol}
		\begin{array}{rclcrclcrcl}
		c_5&=&\frac{1}{4p^2}&;& c_6&=&\frac{1}{4p^4}\left[\frac{4\kappa_3^2}{(\kappa_1 -\kappa_3)^2}-1\right]&;&c_7&=&\frac{1}{2p^2}\\
		c_8&=& -\frac{1}{2p^2}\left[\frac{4\kappa_1 \kappa_3}{(\kappa_1 -\kappa_3)^2}+\frac{\kappa_1+3\kappa_2}{\kappa_1+\kappa_2}\right]&;&c_{10}&=&+\frac{\kappa_1 \kappa_3}{(\kappa_1 -\kappa_3)^2p^2}&;&c_{12}&=&\frac{2}{p^4}\left[\frac{1}{\kappa_1+\kappa_2}-\frac{2}{\kappa_1 -\kappa_3}\right]\\
		c_{14}&=&\frac{2}{(\kappa_1 -\kappa_3)p^2}&;&c_{17}&=&-\frac{\kappa_1^2}{(\kappa_1 -\kappa_3)^2p^2}&;&c_{20}&=&-\frac{2}{(\kappa_1 -\kappa_3)p^2}\ .
		\end{array}
	\ee
The other parameters must vanish: $c_i=0$ for $i=\{0$-$4,9,11,13,15,16,18,19,21$-$23\}$. Defining
	\be
	\bar{\mathcal{E}}_{\alpha\beta,\rho\sigma}\equiv ip^\lambda\left(\epsilon_{\alpha\lambda\rho}p_\sigma p_\beta+\epsilon_{\alpha\lambda\sigma}p_\beta p_\rho+\epsilon_{\beta\lambda\rho}p_\alpha p_\sigma+\epsilon_{\beta\lambda\sigma}p_\alpha p_\rho\right)\ ,
	\ee
taking into account the solutions \eqref{sol} for the coefficients $c_i$, the propagators  \eqref{app-Delta}
-\eqref{app-B} are
	\begin{empheq}{align}
	{H}_{\alpha\beta,\rho\sigma}&=\frac{1}{4}\frac{\mathcal{E}_{\alpha\beta,\rho\sigma}}{p^2}+\frac{1}{4}\left[\frac{4\kappa_3^2}{(\kappa_1 -\kappa_3)^2}-1\right]\frac{\bar{\mathcal{E}}_{\alpha\beta,\rho\sigma}}{p^4}\label{app-Delta'}\\
	{H}_{\alpha\beta,\rho}&= \frac{i}{2}\frac{\eta_{\alpha\beta}p_{\rho}}{p^2} -\frac{i}{2}\left[\frac{4\kappa_1 \kappa_3}{(\kappa_1 -\kappa_3)^2}+\frac{2\kappa_2}{\kappa_1+\kappa_2}+1\right]\frac{p_\alpha p_\beta p_{\rho}}{p^4}+\frac{i\kappa_1 \kappa_3}{(\kappa_1 -\kappa_3)^2}\frac{\eta_{\alpha\rho}p_\beta+\eta_{\beta\rho}p_\alpha}{p^2}\label{app-D'}\\
	{\mathcal{H}}_{\alpha\beta,\rho}&= 2i\left[\frac{1}{\kappa_1+\kappa_2}-\frac{2}{\kappa_1 -\kappa_3}\right]\frac{p_\alpha p_\beta p_{\rho}}{p^4}+\frac{2i}{(\kappa_1 -\kappa_3)}\frac{\eta_{\alpha\rho}p_\beta+\eta_{\beta\rho}p_\alpha}{p^2}\label{app-d'}\\
	A_{\alpha\rho}&= -\frac{i\kappa_1^2}{(\kappa_1 -\kappa_3)^2}\frac{p^\lambda\epsilon_{\alpha\lambda\rho}}{p^2}\label{app-A'}\\
	\mathcal{A}_{\alpha\rho}&=-\frac{2i}{(\kappa_1 -\kappa_3)}\frac{p^\lambda\epsilon_{\alpha\lambda\rho}}{p^2}\label{app-McA'}\\
	B_{\alpha\rho}&= 0\ ,\label{app-B'}
	\end{empheq}
which display poles at the gauge parameters $\kappa_2=-\kappa_1$ and $\kappa_3=\kappa_1$.




\end{document}